\documentclass[preprint,showkeys,superscriptaddress,amsmath,amssymb]{revtex4-1}

\usepackage{times}
\usepackage{graphicx}
\usepackage{dcolumn}
\usepackage{amsmath}
\usepackage{color}
\usepackage[labelsep=space]{caption}
\usepackage{setspace}
\usepackage{amssymb}
\usepackage[version=3]{mhchem}
\usepackage{lineno}

\begin{document}

\renewcommand{\captionfont}{\small}
\newcommand{\NatureFormat}{%
		\renewcommand{\figurename}{\textbf{Figure}}
        \renewcommand{\thefigure}{\textbf{\arabic{figure} $|$}}%
     }
\NatureFormat

\title{Controlling the sign of chromatic dispersion in diffractive optics with dielectric metasurfaces}

\author{Ehsan Arbabi}
\affiliation{T. J. Watson Laboratory of Applied Physics, California Institute of Technology, 1200 E. California Blvd., Pasadena, CA 91125, USA}
\author{Amir Arbabi}
\affiliation{T. J. Watson Laboratory of Applied Physics, California Institute of Technology, 1200 E. California Blvd., Pasadena, CA 91125, USA}
\affiliation{Department of Electrical and Computer Engineering, University of Massachusetts Amherst, 151 Holdsworth Way, Amherst, MA 01003, USA}
\author{Seyedeh Mahsa Kamali}
\affiliation{T. J. Watson Laboratory of Applied Physics, California Institute of Technology, 1200 E. California Blvd., Pasadena, CA 91125, USA}
\author{Yu Horie}
\affiliation{T. J. Watson Laboratory of Applied Physics, California Institute of Technology, 1200 E. California Blvd., Pasadena, CA 91125, USA}
\author{Andrei Faraon}
\email{faraon@caltech.edu}
\affiliation{T. J. Watson Laboratory of Applied Physics, California Institute of Technology, 1200 E. California Blvd., Pasadena, CA 91125, USA}


\begin{abstract}
Diffraction gratings disperse light in a rainbow of colors with the opposite order than refractive prisms, a phenomenon known as negative dispersion. While refractive dispersion can be controlled via material refractive index, diffractive dispersion is fundamentally an interference effect dictated by geometry. Here we show that this fundamental property can be altered using dielectric metasurfaces, and we experimentally demonstrate diffractive gratings and focusing mirrors with positive, zero, and hyper negative dispersion. These optical elements are implemented using a reflective metasurface composed of dielectric nano-posts that provide simultaneous control over phase and its wavelength derivative. In addition, as a first practical application, we demonstrate a focusing mirror that exhibits a five fold reduction in chromatic dispersion, and thus an almost three times increase in operation bandwidth compared to a regular diffractive element. This concept challenges the generally accepted dispersive properties of diffractive optical devices and extends their applications and functionalities.
\end{abstract}

\maketitle

\section{Introduction}
Most optical materials have positive (normal) dispersion, which means that the refractive index decreases at longer wavelengths. As a consequence, blue light is deflected more than red light by dielectric prisms [Fig. \ref{fig:1_concept}(a)]. The reason why diffraction gratings are said to have negative dispersion is because they disperse light similar to hypothetical refractive prisms made of a material with negative (anomalous) dispersion [Fig. \ref{fig:1_concept}(b)]. For diffractive devices, dispersion is not related to material properties, and it refers to the derivative of a certain device parameter with respect to wavelength. For example, the angular dispersion of a grating that deflects normally incident light by a positive angle $\theta$ is given by $\mathrm{d}\theta/\mathrm{d}\lambda=\tan(\theta)/\lambda$ (see~\cite{Born1999} and Supplementary Section S2). Similarly, the wavelength dependence of the focal length ($f$) of a diffractive lens is given by $\mathrm{d}f/\mathrm{d}\lambda=-f/\lambda$ ~\cite{Born1999,O'shea2004}. Here we refer to diffractive devices that follow these fundamental chromatic dispersion relations as ``\textit{regular}". Achieving new regimes of dispersion control in diffractive optics is important both at the fundamental level and for numerous practical applications. Several distinct regimes can be differentiated as follows. Diffractive devices are dispersionless when the derivative is zero (i.e. $\mathrm{d}\theta/\mathrm{d}\lambda=0$, $\mathrm{d}f/\mathrm{d}\lambda=0$ shown schematically in Fig. \ref{fig:1_concept}(c)), have positive dispersion when the derivative has opposite sign compared to a regular diffractive device of the same kind (i.e. $\mathrm{d}\theta/\mathrm{d}\lambda<0$, $\mathrm{d}f/\mathrm{d}\lambda>0$) as shown in Fig. \ref{fig:1_concept}(d), and are hyper-dispersive when the derivative has a larger absolute value than a regular device (i.e. $|\mathrm{d}\theta/\mathrm{d}\lambda|>|\tan(\theta)/\lambda|$, $|\mathrm{d}f/\mathrm{d}\lambda|>|-f/\lambda|$) as seen in Fig. \ref{fig:1_concept}(e). Here we show that these regimes can be achieved in diffractive devices based on optical metasurfaces.

Metasurfaces have attracted great interest in the recent years~\cite{Kildishev2013Science,Yu2014NatMater,Koenderink2015Science,Jahani2016NatNano,Lalanne1998OptLett,Lalanne1999JOSAA,Fattal2010NatPhoton,Yin2013Science,Lee2014Nature,Silva2014Science} because they enable precise control of optical wavefronts and are easy to fabricate with conventional microfabrication technology in a flat, thin, and light weight form factor. Various conventional devices such as gratings, lenses, holograms, and planar filter arrays~\cite{Lalanne1998OptLett,Lalanne1999JOSAA,Fattal2010NatPhoton,Ni2013LightSciApp,Vo2014IEEEPhotonTechLett,Lin2014Science,Arbabi2015NatCommun,Yu2015LaserPhotonRev,Arbabi2015OptExp,Decker2015AdvOptMat,Wang2016NatPhoton,Kamali2016LaserPhotonRev,Zhan2016ACSPhotonics,Arbabi2016NatCommun,Khorasaninejad2016Science,Horie2015OptExp,Horie2016OptExp}, as well as novel devices~\cite{Arbabi2015NatNano,Kamali2016NatCommun} have been demonstrated using metasurfaces. These optical elements are composed of large numbers of scatterers, or meta-atoms placed on a two-dimensional lattice to locally shape optical wavefronts. Similar to other diffractive devices, metasurfaces that locally change the propagation direction (e.g. lenses, beam deflectors, holograms) have negative chromatic dispersion~\cite{Born1999,O'shea2004,Sauvan2004OptLett,Arbabi2016Optica}. This is because most of these devices are divided in Fresnel zones whose boundaries are designed for a specific wavelength~\cite{Faklis1995ApplOpt,Arbabi2016Optica}. This chromatic dispersion is an important limiting factor in many applications and its control is of great interest. Metasurfaces with zero and positive dispersion would be useful for making achromatic singlet and doublet lenses, and the larger-than-regular dispersion of hyper-dispersive metasurface gratings would enable high resolution spectrometers. We emphasize that the devices with zero chromatic dispersion discussed here are fundamentally different from the multiwavelength metasurface gratings and lenses recently reported~\cite{Faklis1995ApplOpt,Eisenbach2015OptExp,Aieta2015Science,Khorasaninejad2015NanoLett,Arbabi2016Optica,Wang2016NanoLett,Arbabi2016OptExp,Zhao2016OptLett,Deng2016OptExp,Arbabi2016SciRep,Lin2016NanoLett}. Multiwavelength devices have several diffraction orders, which result in lenses (gratings) with the same focal length (deflection angle) at a few discrete wavelengths. However, at each of these focal distances (deflection angles), the multi-wavelength lenses (gratings) exhibit the regular negative diffractive chromatic dispersion (see~\cite{Faklis1995ApplOpt,Arbabi2016Optica}, Supplementary Section S3 and Fig. S1).

\section{Theory}
Here we argue that simultaneously controlling the phase imparted by the meta-atoms composing the metasurface ($\phi$) and its derivative with respect to frequency $\omega$ ($\phi'=\partial \phi/\partial\omega$ which we refer to as chromatic phase dispersion or dispersion for brevity) makes it possible to dramatically alter the fundamental chromatic dispersion of diffractive components. This, in effect, is equivalent to simultaneously controlling the ``effective refractive index" and ``chromatic dispersion" of the meta-atoms. We have used this concept to demonstrate metasurface focusing mirrors with zero dispersion~\cite{Arbabi2016CLEO_Displess} in near IR. More recently, the same structure as the one used in~\cite{Arbabi2016CLEO_Displess} (with titanium dioxide replacing $\alpha$:Si) was used to demonstrate achromatic reflecting mirrors in the visible~\cite{Khorasaninejad2017NanoLett}. Using the concept introduced in~\cite{Arbabi2016CLEO_Displess}, here we experimentally show metasurface gratings and focusing mirrors that have positive, zero, and hyper chromatic dispersions. We also demonstrate an achromatic focusing mirror with a highly diminished focal length chromatic dispersion, resulting in an almost three times increase in its operation bandwidth.

First, we consider the case of devices with zero chromatic dispersion. In general for truly frequency independent operation, a device should impart a constant delay for different frequencies (i.e. demonstrate a true time delay behavior), similar to a refractive device made of a non-dispersive material~\cite{Born1999}. Therefore, the phase profile will be proportional to the frequency:
\begin{equation}
\phi(x,y;\omega) = \omega T(x,y),
\label{eq:displess_phase}\\
\end{equation}
where $\omega=2\pi c/ \lambda$ is the angular frequency ($\lambda$: wavelength, $c$: speed of light) and $T(x,y)$ determines the function of the device (for instance $T(x,y)=-x \sin{\theta_0}/c$ for a grating that deflects light by angle $\theta_0$; $T(x,y)=-\sqrt{x^2 +y^2 + f^2}/c$ for a spherical-aberration-free lens with a focal distance $f$). Since the phase profile is a linear function of $\omega$, it can be realized using a metasurface composed of meta-atoms that control the phase $\phi(x,y;\omega_0) = T(x,y)\omega_0$ and its dispersion $\phi'=\mathrm{\partial}\phi(x,y;\omega)/\mathrm{\partial}\omega = T(x,y)$. The bandwidth of dispersionless operation corresponds to the frequency interval over which the phase locally imposed by the meta-atoms is linear with frequency $\omega$. For gratings or lenses, a large device size results in a large $|T(x,y)|$, which means that the meta-atoms should impart a large phase dispersion. Since the phase values at the center wavelength $\lambda_0=2\pi c/\omega_0$ can be wrapped into the 0 to 2$\pi$ interval, the meta-atoms only need to cover a rectangular region in the \textit{phase-dispersion} plane bounded by $\phi=0$ and 2$\pi$ lines, and $\phi'=0$ and $\phi'_\mathrm{max}$ lines, where $\phi'_\mathrm{max}$ is the maximum required dispersion which is related to the device size (see Supplementary Section S5 and Fig. S2). The required phase-dispersion coverage means that, to implement devices with various phase profiles, for each specific value of the phase we need various meta-atoms providing that specific phase, but with different dispersion values.

Considering the simple case of a flat dispersionless lens (or focusing mirror) with radius $R$, we can get some intuition to the relations found for phase and dispersion. Dispersionless operation over a certain bandwidth $\Delta \omega$ means that the device should be able to focus a transform limited pulse with bandwidth $\Delta \omega$ and carrier frequency $\omega_0$ to a single spot located at focal length $f$ [Fig. \ref{fig:2_Simulations}(a)]. To implement this device, part of the pulse hitting the lens at a distance $r$ from its center needs to experience a pulse delay (i.e. group delay $t_g = \partial\phi/\partial\omega$) smaller by $(\sqrt{r^2+f^2}-f)/c$ than part of the pulse hitting the lens at its center. This ensures that parts of the pulse hitting the lens at different locations arrive at the focus at the same time. Also, the carrier delay (i.e. phase delay $t_p = \phi(\omega_0)/\omega_0$) should also be adjusted so that all parts of the pulse interfere constructively at the focus. Thus, to implement this phase delay and group delay behavior, the lens needs to be composed of elements, ideally with sub-wavelength size, that can provide the required phase delay and group delay at different locations. For a focusing mirror, these elements can take the form of sub-wavelength one-sided resonators, where the group delay is related to the quality factor $Q$ of the resonator (see Supplementary Section S7) and the phase delay depends on the resonance frequency. We note that larger group delays are required for lenses with larger radius, which means that elements with higher quality factors are needed. If the resonators are single mode, the $Q$ imposes an upper bound on the maximum bandwidth $\Delta \omega$ of the pulse that needs to be focused. The operation bandwidth can be expanded by using one-sided resonators with multiple resonances that partially overlap. As we will show later in the paper, these resonators can be implemented using silicon nano-posts backed by a reflective mirror.

To realize metasurface devices with non-zero dispersion of a certain parameter $\xi(\omega)$, phase profiles of the following form are needed:
\begin{equation}
\phi(x,y;\omega) = \omega T(x,y,\xi(\omega)).
\label{eq:disp_phase}\\
\end{equation}
\noindent{For instance, the parameter $\xi(\omega)$ can be the deflection angle of a diffraction grating $\theta(\omega)$ or the focal length of a diffractive lens $f(\omega)$. As we show in the Supplementary Section S4, to independently control the parameter $\xi(\omega)$ and its chromatic dispersion $\mathrm{\partial}\xi/\mathrm{\partial}\omega$ at $\omega=\omega_0$, we need to control the phase dispersion at this frequency in addition to the phase. The required dispersion for a certain parameter value $\xi_0=\xi(\omega_0)$, and a certain dispersion $\mathrm{\partial}\xi/\mathrm{\partial}\omega|_{\omega=\omega_0}$ is given by:
\begin{equation}
\frac{\mathrm{\partial}\phi(x,y;\omega)}{\mathrm{\partial}\omega}|_{\omega=\omega_0} = T(x,y,\xi_0)+\partial\xi/\partial\omega|_{\omega=\omega_0}\omega_0\frac{\partial T(x,y,\xi)}{\partial\xi}|_{\xi=\xi_0}.
\label{eq:disp_disp_phase}\\
\end{equation}
\noindent{This dispersion relation is valid over a bandwidth where a linear approximation of $\xi(\omega)$ is valid. One can also use Fermat's principle to get similar results to Eq. \ref{eq:disp_disp_phase} for the local phase gradient and its frequency derivative (see Supplementary Section S6).}

We note that discussing these types of devices in terms of phase $\phi(\omega)$ and phase dispersion $\partial\phi/\partial\omega$, which we mainly use in this paper, is equivalent to using the terminology of phase delay ($t_p=\phi(\omega_0)/\omega_0$) and group delay ($t_g=\partial\phi/\partial\omega$). The zero dispersion case discussed above corresponds to a case where the phase and group delays are equal. Figures~\ref{fig:2_Simulations}(b) and \ref{fig:2_Simulations}(c) show the required phase and group delays for blazed gratings and focusing mirrors with various types of dispersion, demonstrating the equality of phase and group delays in the dispersionless case. In microwave photonics, the idea of using sets of separate optical cavities for independent control of the phase delay of the optical carrier, and group delay of the modulated RF signal has previously been proposed~\cite{Morton2009IEEEPhotonTechLett} to achieve dispersionless beam steering and resemble a true time delay system over a narrow bandwidth. For all other types of chromatic dispersion, the phase and group delays are drastically different as shown in Figs.~\ref{fig:2_Simulations}(b) and \ref{fig:2_Simulations}(c).

Assuming hypothetical meta-atoms that provide independent control of phase and dispersion up to a dispersion of $-150$~Rad$/\mu$m (to adhere to the commonly used convention, we report the dispersion in terms of wavelength) at the center wavelength of 1520~nm, we have designed and simulated four gratings with different chromatic dispersions (see Supplementary Section S1 for details). The simulated deflection angles as functions of wavelength are plotted in Fig. \ref{fig:2_Simulations}(d). All gratings are 150~$\mu$m wide, and have a deflection angle of 10 degrees at their center wavelength of 1520~nm. The positive dispersion grating exhibits a dispersion equal in absolute value to the negative dispersion of a regular grating with the same deflection angle, but with an opposite sign. The hyper-dispersive design is three times more dispersive than the regular grating, and the dispersionless beam deflector shows almost no change in its deflection angle. Besides gratings, we have also designed focusing mirrors exhibiting regular, zero, positive, and hyper dispersions. The focusing mirrors have a diameter of $500$~$\mu$m and a focal distance of $850$~$\mu$m at 1520~nm. Hypothetical meta-atoms with a maximum dispersion of $-200$~Rad$/\mu$m are required to implement these focusing mirror designs. The simulated focal distances of the four designs are plotted in Fig. \ref{fig:2_Simulations}(e). The axial plane intensity distributions at three wavelengths are plotted in Figs. \ref{fig:2_Simulations}(f-i) (for intensity plots at other wavelengths see Supplementary Fig. S3). To relate to our previous discussion of dispersionless focusing mirrors depicted in Fig. \ref{fig:2_Simulations}(a), a focusing mirror with a diameter of 500~$\mu$m and a focal distance of 850~$\mu$m would require meta-atoms with group delay of $\sim$24~$\lambda_0/\mathrm{c}$, with $\lambda_0$=1520~nm. To implement this device we used hypothetical meta-atoms with maximum dispersion of $\sim$-100~Rad/$\mu$m which corresponds to a group delay of $\sim$24~$\lambda_0/\mathrm{c}$. The hypothetical meta-atoms exhibit this almost linear dispersion over the operation bandwidth of 1450~nm to 1590~nm.

\section{Metasurface design}
An example of meta-atoms capable of providing 0 to 2$\pi$ phase coverage and different dispersions is shown in Fig. \ref{fig:3_DesignGraphs}(a). The meta-atoms are composed of a square cross-section amorphous silicon ($\mathrm{\alpha}$-Si) nano-post on a low refractive index silicon dioxide (SiO$_2$) spacer layer on an aluminum reflector that play the role of the multi-mode one sided resonators mentioned in Section 2 [Fig. \ref{fig:2_Simulations}(a)]. They are located on a periodic square lattice [Fig. \ref{fig:3_DesignGraphs}(a), middle]. The simulated dispersion versus phase plot for the meta-atoms at the wavelength of $\lambda_0=1520$~nm is depicted in Fig. \ref{fig:3_DesignGraphs}(b), and shows a partial coverage up to the dispersion value of $\sim-100$~Rad$/\mu$m. The nano-posts exhibit several resonances which enables high dispersion values over the 145~0nm to 1590~nm wavelength range. The meta-atoms are 725~nm tall, the SiO$_{2}$ layer is 325~nm thick, the lattice constant is 740~nm, and the nano-post side length is varied from 74 to 666~nm at 1.5~nm steps. Simulated reflection amplitude and phase for the periodic lattice are plotted in Figs. \ref{fig:3_DesignGraphs}(c) and \ref{fig:3_DesignGraphs}(d), respectively. The reflection amplitude over the bandwidth of interest is close to 1 for all nano-post side lengths. The operation of the nano-post meta-atoms is best intuitively understood as truncated multi-mode waveguides with many resonances in the bandwidth of interest \cite{Kamali2016NatCommun,Lalanne1999JOSAA_Multimode}. By going through the nano-post twice, light can obtain larger phase shifts compared to the transmissive operation mode of the metasurface (i.e. without the metallic reflector). The metallic reflector keeps the reflection amplitude high for all sizes, which makes the use of high quality factor resonances possible. As discussed in Section 2, high quality factor resonances are necessary for achieving large dispersion values, because, as we have shown in Supplementary Section S7, dispersion is given by $\phi'\approx -Q/\lambda_0$, where $Q$ is the quality factor of the resonance.

Using the dispersion-phase parameters provided by this metasurface, we designed four gratings operating in various dispersion regimes. The gratings are $\sim$90~$\mu$m wide and have a 10-degree deflection angle at 1520~nm. They are designed to operate in the 1450 to 1590~nm wavelength range, and have regular negative, zero, positive, and hyper (three-times-larger negative) dispersion. Since the phase of the meta-atoms does not follow a linear frequency dependence over this wavelength interval [Fig. \ref{fig:3_DesignGraphs}(d), top right], we calculate the desired phase profile of the devices at 8 wavelengths in the range (1450 to 1590~nm at 20~nm steps), and form an 8$\times$1 complex reflection coefficient vector at each point on the metasurface. Using Figs. \ref{fig:3_DesignGraphs}(c) and \ref{fig:3_DesignGraphs}(d), a similar complex reflection coefficient vector is calculated for each meta-atom. Then, at each lattice site of the metasurface, we place a meta-atom whose reflection vector has the shortest weighted Euclidean distance to the desired reflection vector at that site. The weights allow for emphasizing different parts of the operation bandwidth, and can be chosen based on the optical spectrum of interest or other considerations. Here, we used an inverted Gaussian weight ($\exp((\lambda-\lambda_0)^2/2\sigma^2)$, $\sigma=300$~nm), which values wavelengths farther away from the center wavelength of $\lambda_0=1520$~nm. The same design method is used for the other devices discussed in the manuscript. The designed devices were fabricated using standard semiconductor fabrication techniques as described in Supplementary Section S1. Figures \ref{fig:3_DesignGraphs}(e) and \ref{fig:3_DesignGraphs}(f) show scanning electron micrographs of the nano-posts, and some of the devices fabricated using the proposed reflective meta-atoms. Supplementary Figure S5 shows the chosen post side lengths and the required as well as the achieved phase and group delays for the gratings with different dispersions. Required phases, and the values provided by the chosen nano-posts are plotted at three wavelengths for each grating in Supplementary Fig. S6.

\section{Experimental results}
Figures \ref{fig:4_Gratings}(a) and \ref{fig:4_Gratings}(b) show the simulated and measured deflection angles for gratings, respectively. The measured values are calculated by finding the center of mass of the deflected beam 3~mm away from the grating surface (see Supplementary Section S1 and Fig. S8 for more details). As expected, the zero dispersion grating shows an apochromatic behavior resulting in a reduced dispersion, the positive grating shows positive dispersion in the $\sim$1490-1550~nm bandwidth, and the hyper-dispersive one shows an enhanced dispersion in the measurement bandwidth. This can also be viewed from the grating momentum point of view: a regular grating has a constant momentum set by its period, resulting in a constant transverse wave-vector. In contrary, the momentum of the hyper-dispersive grating increases with wavelength, while that of the zero and positive gratings decreases with it. This means that the effective period of the non-regular gratings changes with wavelength, resulting in the desired chromatic dispersion. Figures \ref{fig:4_Gratings}(e-h) show good agreement between simulated intensities of these gratings versus wavelength and transverse wave-vector (see Supplementary Section S1 for details) and the measured beam deflection (black stars). The change in the grating pitch with wavelength is more clear in Supplementary Fig. S6, where the required and achieved phases are plotted for three wavelengths. The green line is the theoretical expectation of the maximum intensity trajectory. Measured deflection efficiencies of the gratings, defined as the power deflected by the gratings to the desired order, divided by the power reflected from a plain aluminum reflector (see Supplementary Section S1 and Fig. S8 for more details) are plotted in Figs. \ref{fig:4_Gratings}(c) and \ref{fig:4_Gratings}(d) for TE and TM illuminations, respectively. A similar difference in the efficiency of the gratings for TE and TM illuminations has also been observed in previous works~\cite{Arbabi2015NatCommun,Kamali2016NatCommun}.

As another example for diffractive devices with controlled chromatic dispersion, four spherical-aberration-free focusing mirrors with different chromatic dispersions were designed, fabricated and measured using the same reflective dielectric meta-atoms. The mirrors are 240~$\mu$m in diameter and are designed to have a focal distance of 650~$\mu$m at 1520~nm. Supplementary Figure S6 shows the chosen post side lengths and the required as well as the achieved phase and group delays for the focusing mirrors with different dispersions. Figures \ref{fig:5_Lenses}(a) and \ref{fig:5_Lenses}(b) show simulated and measured focal distances for the four focusing mirrors (see Supplementary Figs. S9, S10, and S11 for detailed simulation and measurement results). The positive dispersion mirror is designed with dispersion twice as large as a regular mirror with the same focal distance, and the hyper-dispersive mirror has a negative dispersion three and a half times larger than a regular one. The zero dispersion mirror shows a significantly reduced dispersion, while the hyper-dispersive one shows a highly enhanced dispersion. The positive mirror shows the expected dispersion in the $\sim$1470 to 1560~nm range.

As an application of diffractive devices with dispersion control, we demonstrate a spherical-aberration-free focusing mirror with increased operation bandwidth. For brevity, we call this device dispersionless mirror. Since the absolute focal distance change is proportional to the focal distance itself, a relatively long focal distance is helpful for unambiguously observing the change in the device dispersion. Also, a higher NA value is preferred because it results in a shorter depth of focus, thus making the measurements easier. Having these considerations in mind, we have chosen a diameter of 500~$\mu$m and a focal distance of 850~$\mu$m (NA$\approx$0.28) for the mirror, requiring a maximum dispersion of $\phi'_\mathrm{max}\approx-98$~Rad/$\mu$m which is achievable with the proposed reflective meta-atoms. We designed two dispersionless mirrors with two $\sigma$ values of 300 and 50~nm. For comparison, we also designed a regular metasurface mirror for operation at $\lambda_0=1520$~nm and with the same diameter and focal distance as the dispersionless mirrors. The simulated focal distance deviations (from the designed 850~$\mu$m) for the regular and dispersionless ($\sigma=300$~nm) mirrors are plotted in Fig. \ref{fig:5_Lenses}(c), showing a considerable reduction in chromatic dispersion for the dispersionless mirror. Detailed simulation results for these mirrors are plotted in Supplementary Fig. S12.

Figures \ref{fig:5_Lenses}(d-g) summarize the measurement results for the dispersionless and regular mirrors (see Supplementary Section S1 and Fig. S8 for measurement details and setup). As Figs. \ref{fig:5_Lenses}(d) and \ref{fig:5_Lenses}(g) show, the focal distance of the regular mirror changes almost linearly with wavelength. The dispersionless mirror, however, shows a highly diminished chromatic dispersion. Besides, as seen from the focal plane intensity measurements, while the dispersionless mirrors are in focus in the 850~$\mu$m plane throughout the measured bandwidth, the regular mirror is in focus only from 1500 to 1550~nm (see Supplementary Figs. S13 and S14 for complete measurement results, and the Strehl ratios). Focusing efficiencies, defined as the ratio of the optical power focused by the mirrors to the power incident on them, were measured at different wavelengths for the regular and dispersionless mirrors (see Supplementary Section S1 for details). The measured efficiencies were normalized to the efficiency of the regular metasurface mirror at its center wavelength of 1520~nm (which is estimated to be $\sim$80$\%$--90$\%$ based on Fig. \ref{fig:3_DesignGraphs}, measured grating efficiencies, and our previous works~\cite{Arbabi2015NatCommun}).  The normalized efficiency of the dispersionless mirror is between 50$\%$ and 60$\%$ in the whole wavelength range and shows no significant reduction in contrast to the regular metasurface mirror.

\section{Discussion and conclusion}
The reduction in efficiency compared to a mirror designed only for the center wavelength (i.e. the regular mirror) is caused by two main factors. First, the required region of the phase-dispersion plane is not completely covered by the reflective nano-post meta-atoms. Second, the meta-atom phase does not change linearly with respect to frequency in the relatively large bandwidth of 140~nm as would be ideal for a dispersionless metasurface. Both of these factors result in deviation of the phase profiles of the demonstrated dispersionless mirrors from the ideal ones. Furthermore, dispersionless metasurfaces use meta-atoms supporting resonances with high quality factors, thus leading to higher sensitivity of these devices to fabrication errors compared to the regular metasurfaces.

Equation \ref{eq:disp_disp_phase} is basically a Taylor expansion of Eq. \ref{eq:disp_phase} kept to the first order. As a result, this equation is accurate only over the range of linearity of the phase given in Eq. \ref{eq:disp_phase}. To increase the validity bandwidth, one can generalize the method to keep higher order terms of the series. Another method to address this issue is the Euclidean distance minimization method that was used in the design process of the devices presented here.

In conclusion, we demonstrated that independent control over phase and dispersion of meta-atoms can be used to engineer the chromatic dispersion of diffractive metasurface devices over continuous wavelength regions. This is in effect similar to controlling the ``material dispersion" of meta-atoms to compensate, over-compensate, or increase the structural dispersion of diffractive devices. In addition, we developed a reflective dielectric metasurface platform that provides this independent control. Using this platform, we experimentally demonstrated gratings and focusing mirrors exhibiting positive, negative, zero, and enhanced dispersions. We also corrected the chromatic aberrations of a focusing mirror resulting in a $\sim$3 times bandwidth increase (based on an Strehl ratio $>0.6$, see Supplementary Fig. S14). In addition, the introduced concept of metasurface design based on dispersion-phase parameters of the meta-atoms is general and can also be used for developing transmissive dispersion engineered metasurface devices.


\vspace{0.2in}
\section*{Acknowledgements}
This work was supported by Samsung Electronics. E.A. and A.A. were also supported by National Science Foundation award 1512266. A.A. and Y.H. were also supported by DARPA, and S.M.K. was supported as part of the Department of Energy (DOE) ``Light-Material Interactions in Energy Conversion‚" Energy Frontier Research Center under grant no. DE-SC0001293. The device nanofabrication was performed at the Kavli Nanoscience Institute at Caltech.

\noindent\textbf{Author contributions} 
E.A., A.A., and A.F. conceived the experiment. E.A., S.M.K., and Y.H. fabricated the samples. E.A., S.M.K., A.A., and Y.H. performed the simulations, measurements, and analyzed the data. E.A., A.F., and A.A. co-wrote the manuscript. All authors discussed the results and commented on the manuscript.


\clearpage

\begin{figure*}[htp]
\centering
\includegraphics{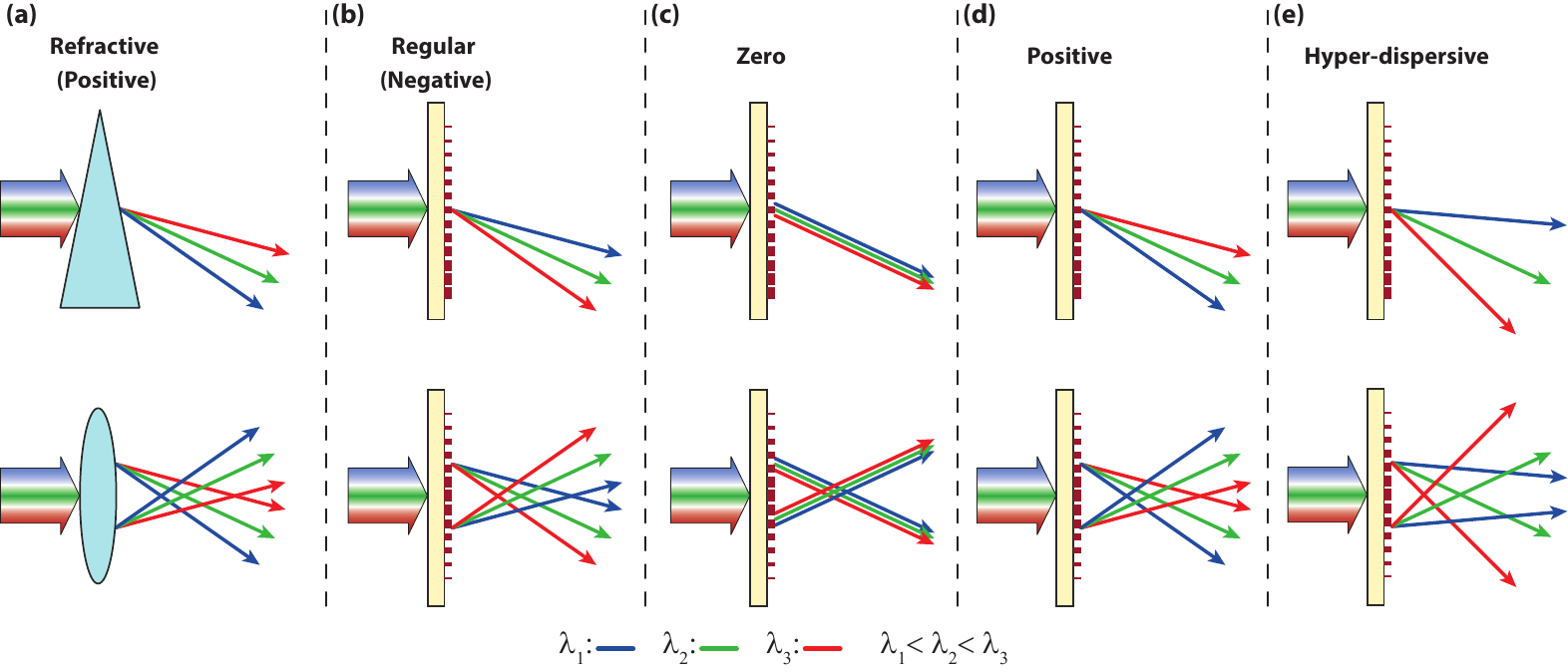}
\caption{Schematic illustrations of different dispersion regimes. (a) Positive chromatic dispersion in refractive prisms and lenses made of materials with normal dispersion. (b) Regular (negative) dispersion in typical diffractive and metasurface gratings and lenses. (c) Schematic illustration of zero, (d) positive, and (e) hyper dispersion in dispersion-controlled metasurfaces. Only three wavelengths are shown here, but the dispersions are valid for any other wavelength in the bandwidth. The diffractive devices are shown in transmission mode for ease of illustration, while the actual devices fabricated in this paper are designed to operate in reflection mode.}
\label{fig:1_concept}
\end{figure*}
\clearpage

\begin{figure*}[htp]
\centering
\includegraphics[width=1\columnwidth]{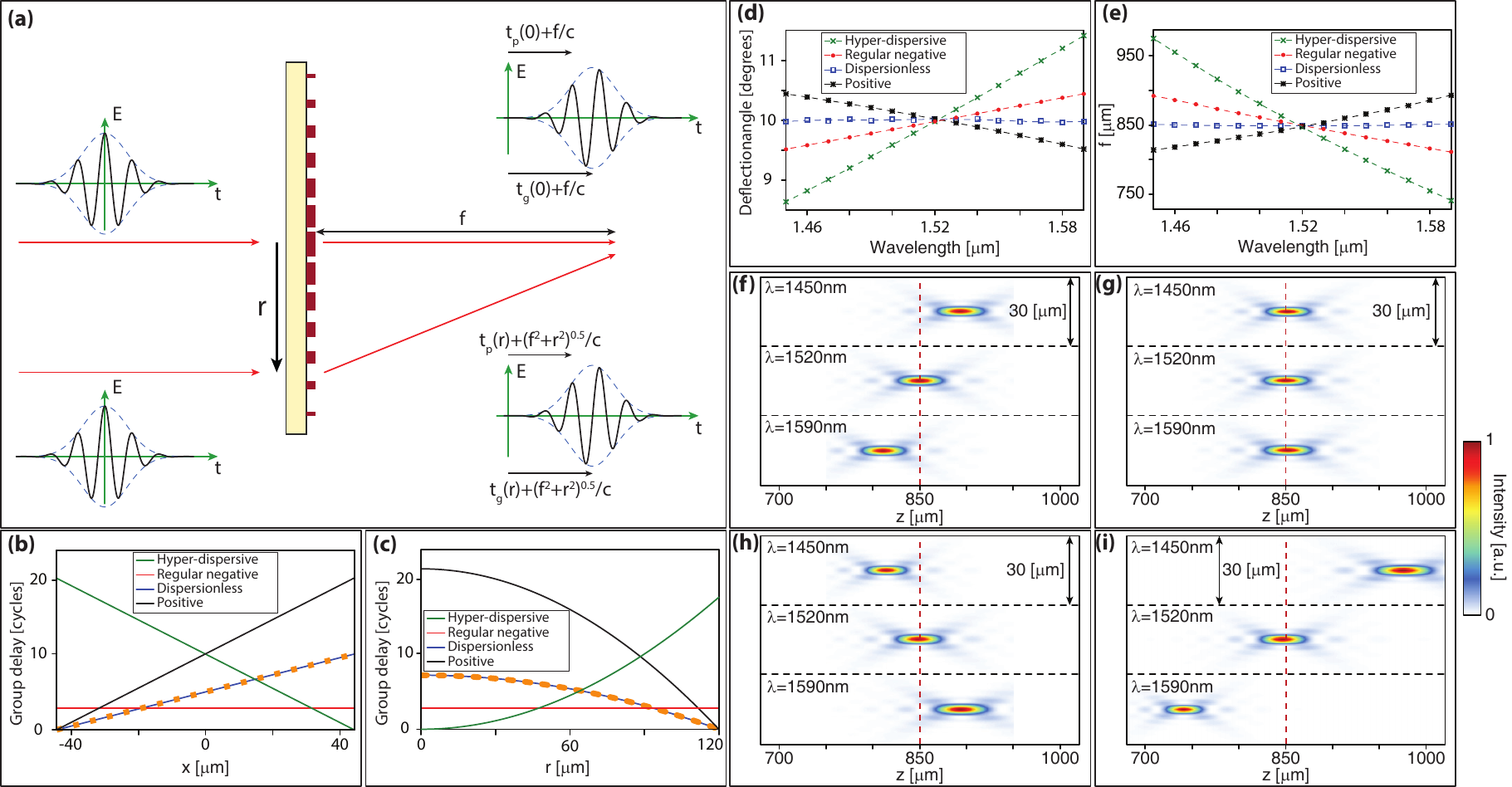}
\caption{Required phase and group delays and simulation results of dispersion-engineered metasurfaces based on hypothetical meta-atoms. (a) Schematics of focusing of a light pulse to the focal distance of a flat lens. The $E$ vs $t$ graphs show schematically the portions of the pulse passing through the center and at a point at a distance $r$ away from center both before the lens, and when arriving at focus. The portions passing through different parts of the lens should acquire equal group delays, and should arrive at the focal point in phase for dispersionless operation.(b) Required values of group delay for gratings with various types of chromatic dispersion. The dashed line shows the required phase delay for all devices, which also coincides with the required group delay for the dispersionless gratings. The gratings are $\sim$90~$\mu$m wide, and have a deflection angle of 10 degrees in their center wavelength of 1520~nm. (c) Required values of group delay for aspherical focusing mirrors with various types of chromatic dispersion. The dashed line shows the required phase delay for all devices. The mirrors are 240~$\mu$m in diameter, and have a focal distance of 650~$\mu$m at their center wavelength of 1520~nm. (d) Simulated deflection angles for gratings with regular, zero, positive, and hyper dispersions. The gratings are 150~$\mu$m wide and have a 10-degree deflection angle at 1520~nm. (e) Simulated focal distances for metasurface focusing mirrors with different types of dispersion. The mirrors are 500~$\mu$m in diameter and have a focal distance of 850~$\mu$m at 1520~nm. All gratings and focusing mirrors are designed using hypothetical meta-atoms that provide independent control over phase and dispersion (see Supplementary Section S1 for details). (f) Intensity in the axial plane for the focusing mirrors with regular negative, (g) zero, (h) positive, and (i) hyper dispersions plotted at three wavelengths (see Supplementary Fig. S3 for other wavelengths).}
\label{fig:2_Simulations}
\end{figure*}
\clearpage

\begin{figure}[htp]
\centering
\includegraphics[width=0.6\columnwidth]{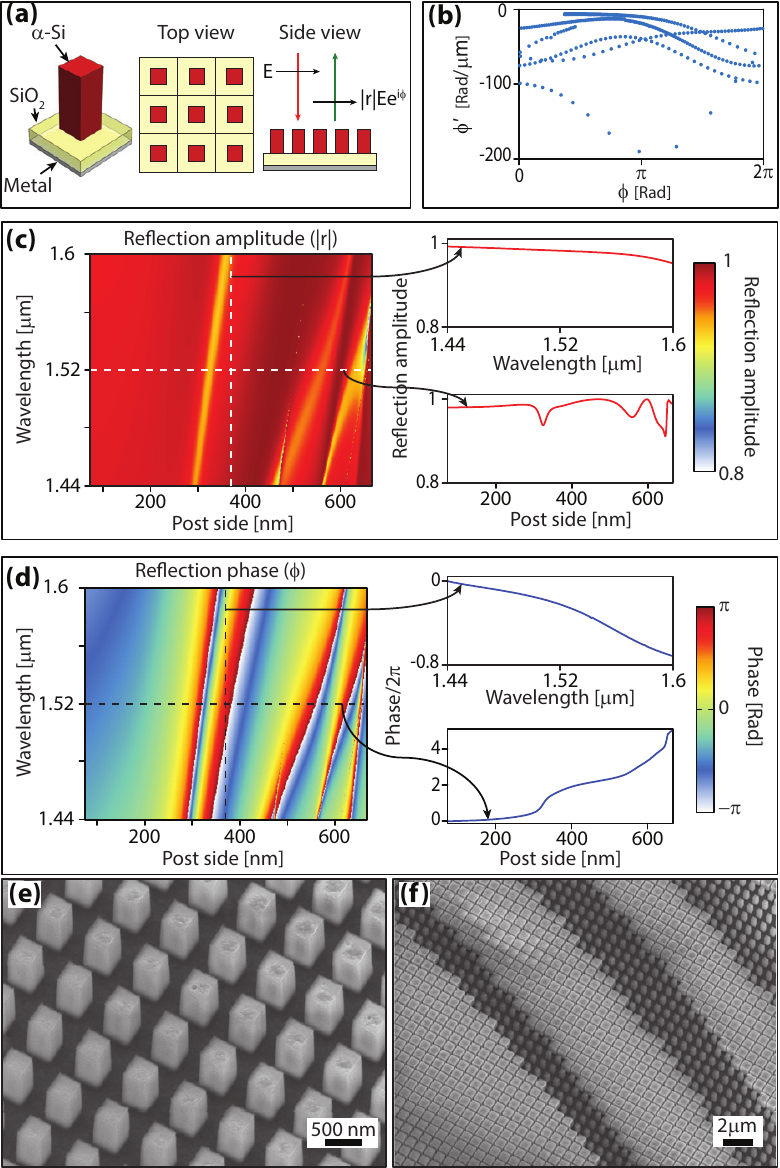}
\caption{High dispersion silicon meta-atoms. (a) A meta-atom composed of a square cross-section amorphous silicon nano-post on a silicon dioxide layer on a metallic reflector. Top and side views of the meta-atoms arranged on a square lattice are also shown. (b) Simulated dispersion versus phase plot for the meta-atom shown in (a) at $\lambda_0=$1520~nm. (c) Simulated reflection amplitude, and (d) phase as a function of the nano-post side length and wavelength. The reflection amplitude and phase along the dashed lines are plotted on the right. (e, f) Scanning electron micrographs of the fabricated nano-posts and devices.}
\label{fig:3_DesignGraphs}
\end{figure}
\clearpage

\begin{figure}[htp]
\centering
\includegraphics[width=1\columnwidth]{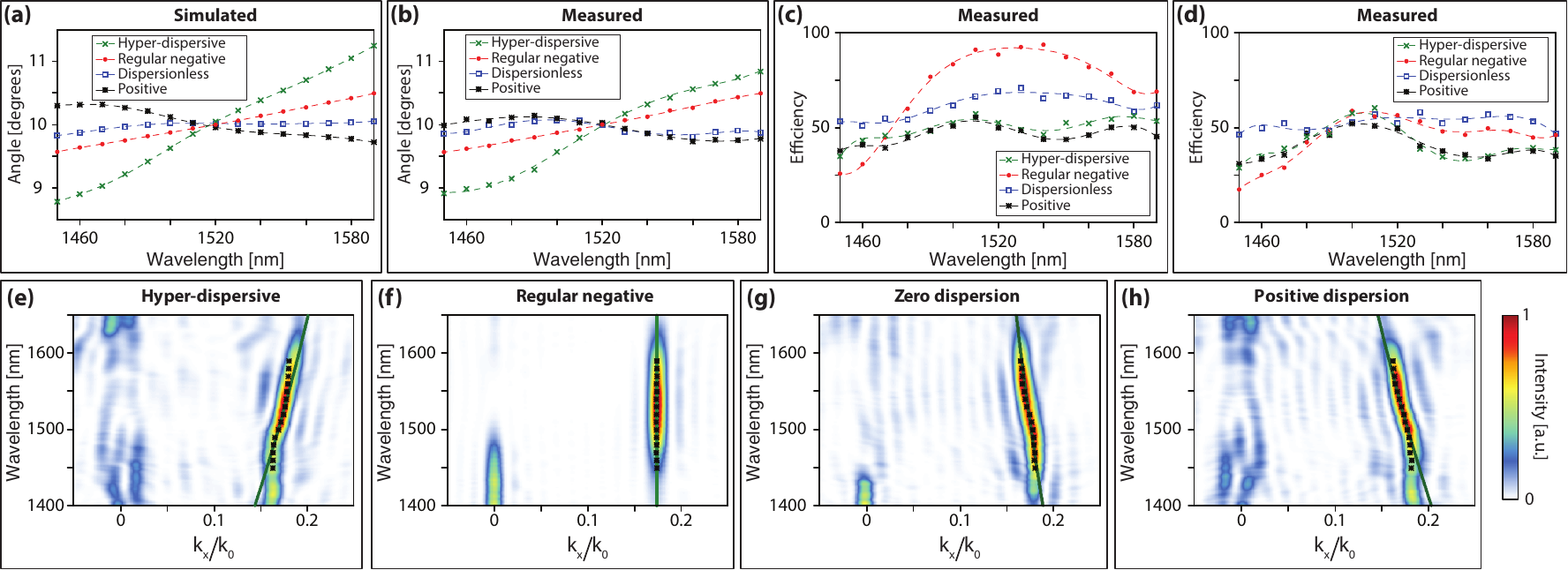}
\caption{Simulation and measurement results of gratings in different dispersion regimes. (a) Simulated deflection angles for gratings with different dispersions, designed using the proposed reflective meta-atoms. (b) Measured deflection angles for the same grating. (c) Measured deflection efficiency for the gratings under TE, and (d) TM illumination. (e-h) Comparison between FDTD simulation results showing the intensity distribution of the diffracted wave as a function of normalized transverse wave-vector ($k_x/k_0$, $k_0=2\pi/\lambda_0$, and $\lambda_0=$1520~nm) and wavelength for different gratings, and the measured peak intensity positions plotted with black stars. All simulations here are performed with TE illumination. The green lines show the theoretically expected maximum intensity trajectories.}
\label{fig:4_Gratings}
\end{figure}
\clearpage

\begin{figure*}[htp]
\centering
\includegraphics[width=1\columnwidth]{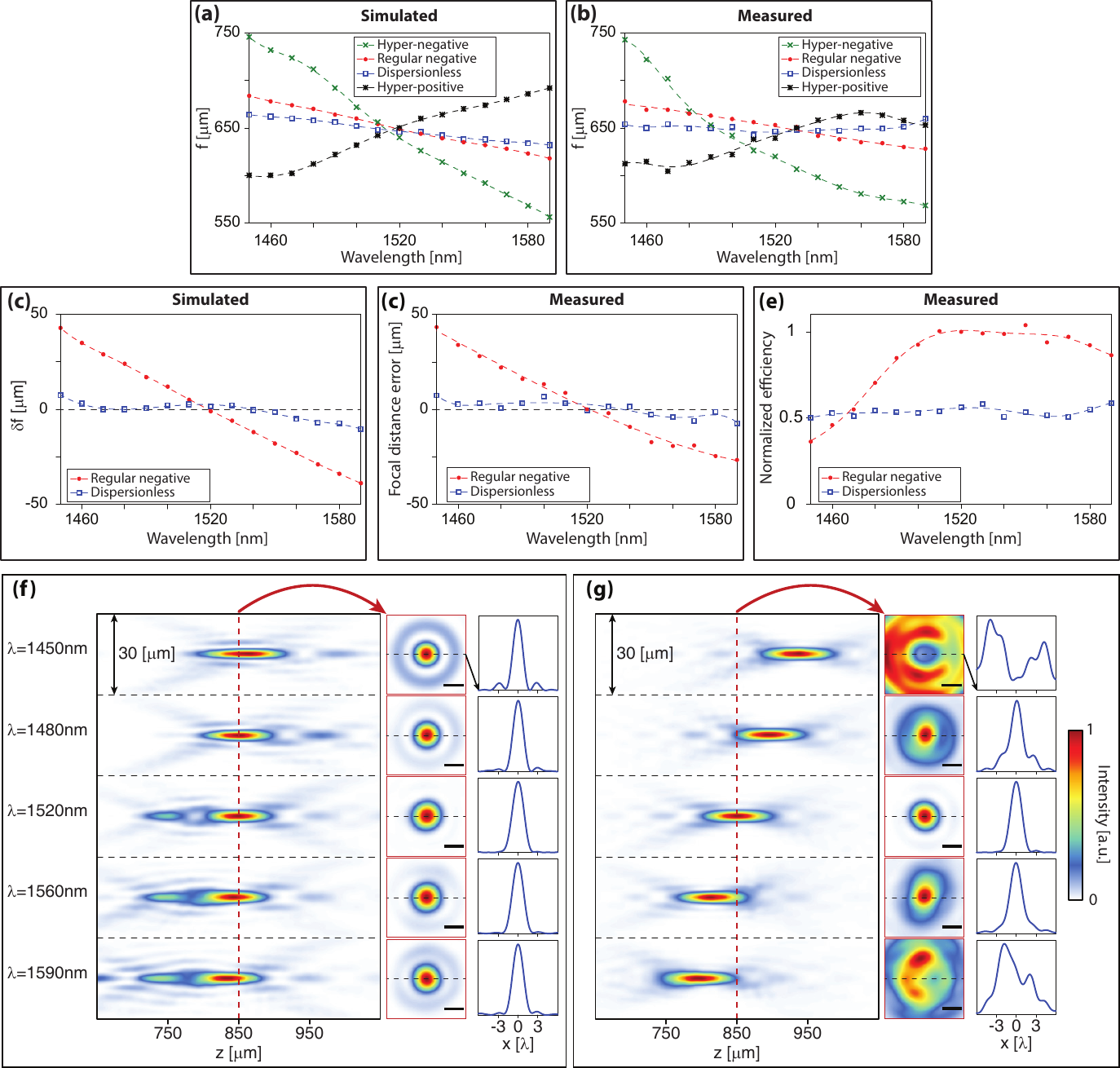}
\caption{Simulation and measurement results for mirrors with different dispersion regimes. (a) Simulated focal distance for focusing mirrors with different dispersions, designed using the reflective meta-atoms (see Supplementary Fig. S9 for axial plane intensity distributions). The mirrors are 240~$\mu$m in diameter and have a focal distance of 650~$\mu$m at 1520~nm. (b) Measured focal distances of the same focusing mirrors (see Supplementary Figs. S10 and S11 for axial plane intensity distributions). (c) Simulated and (d) measured focal distance deviation from its design value of 850~$\mu$m as a function of wavelength for the dispersionless and regular mirrors (see Supplementary Figs. S12 and S13 for extended simulation and measurement results). (e) Measured efficiency for the regular and dispersionless mirrors normalized to the efficiency of the regular device at its center wavelength of 1520~nm. (f) Measured intensity in the axial plane of the dispersionless metasurface mirror at five wavelengths (left). Intensity distributions measured in the desired focal plane (i.e. 850~$\mu$m away from the mirror surface) at the same wavelengths are shown in the center, and their one dimensional profiles along the $x$ axis are plotted on the right. (g) Same plots as in (f) but for the regular mirror. Scale bars: 2$\lambda$.}
\label{fig:5_Lenses}
\end{figure*}
\clearpage

\newcommand{\NatureFormatExtendedData}{%
        \setcounter{figure}{0}
		\renewcommand{\figurename}{\textbf{Extended Data Figure}}
        \renewcommand{\thefigure}{\textbf{\arabic{figure} $|$}}%
     }
\NatureFormatExtendedData

\begin{figure*}[htp]
\centering
\includegraphics[width=1\columnwidth]{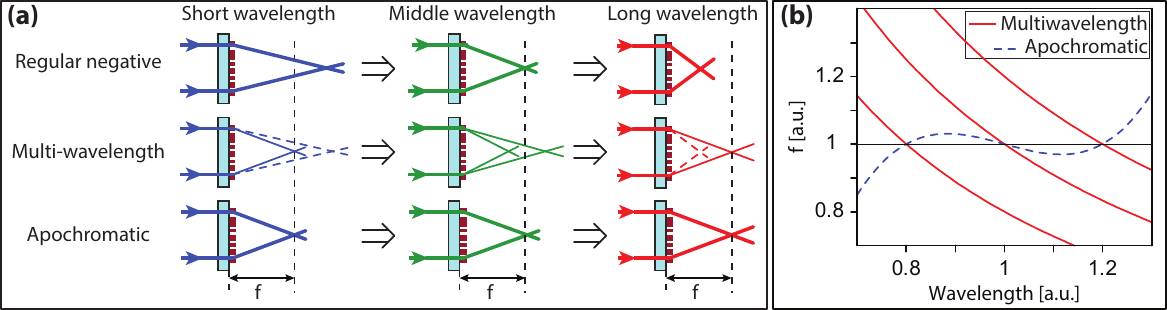}
\caption{Comparison of regular, multi-wavelength, and apochromatic lenses. (a) Schematic comparison of a regular, a multi-wavelength, and an apochromatic metasurface lens. The multi-wavelength lens is corrected at a short and a long wavelength to have a single focal point at a distance $f$, but it has two focal points at wavelengths in between them, none of which is at $f$. The apochromatic lens is corrected at the same short and long wavelengths, and in wavelengths between them it will have a single focus very close to $f$. (b) Focal distances for three focal points of a multiwavelength lens corrected at three wavelengths, showing the regular dispersion (i.e. $f\propto 1/\lambda$) of each focus with wavelength. For comparison, focal distance for the single focus of a typical apochromatic lens is plotted.}
\label{fig:S1_Multi_vs_Ach}
\end{figure*}


\begin{figure*}[htp]
\centering
\includegraphics[width=1\columnwidth]{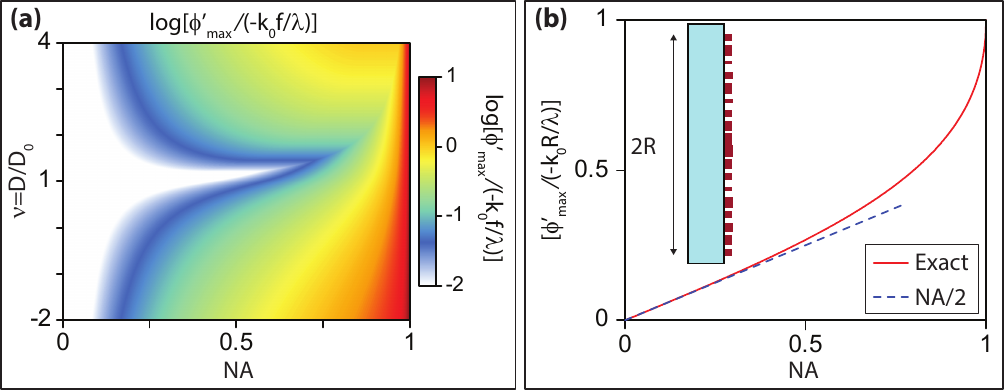}
\caption{Maximum required dispersion of meta-atoms for lenses. (a) Maximum meta-atom dispersion necessary to control the dispersion of a spherical-aberration-free lens. The maximum dispersion is normalized to $-k_0 f/\lambda_0$ and is plotted on a logarithmic scale. (b) Normalized (to $-k_0 R/\lambda_0$) maximum dispersion required for a dispersionless lens. $R$ is the radius, $f$ is the focal distance, and NA is the numerical aperture of the lens.}
\label{fig:S2_MaxDisps}
\end{figure*}

\clearpage

\begin{figure*}[htp]
\centering
\includegraphics[width=1\columnwidth]{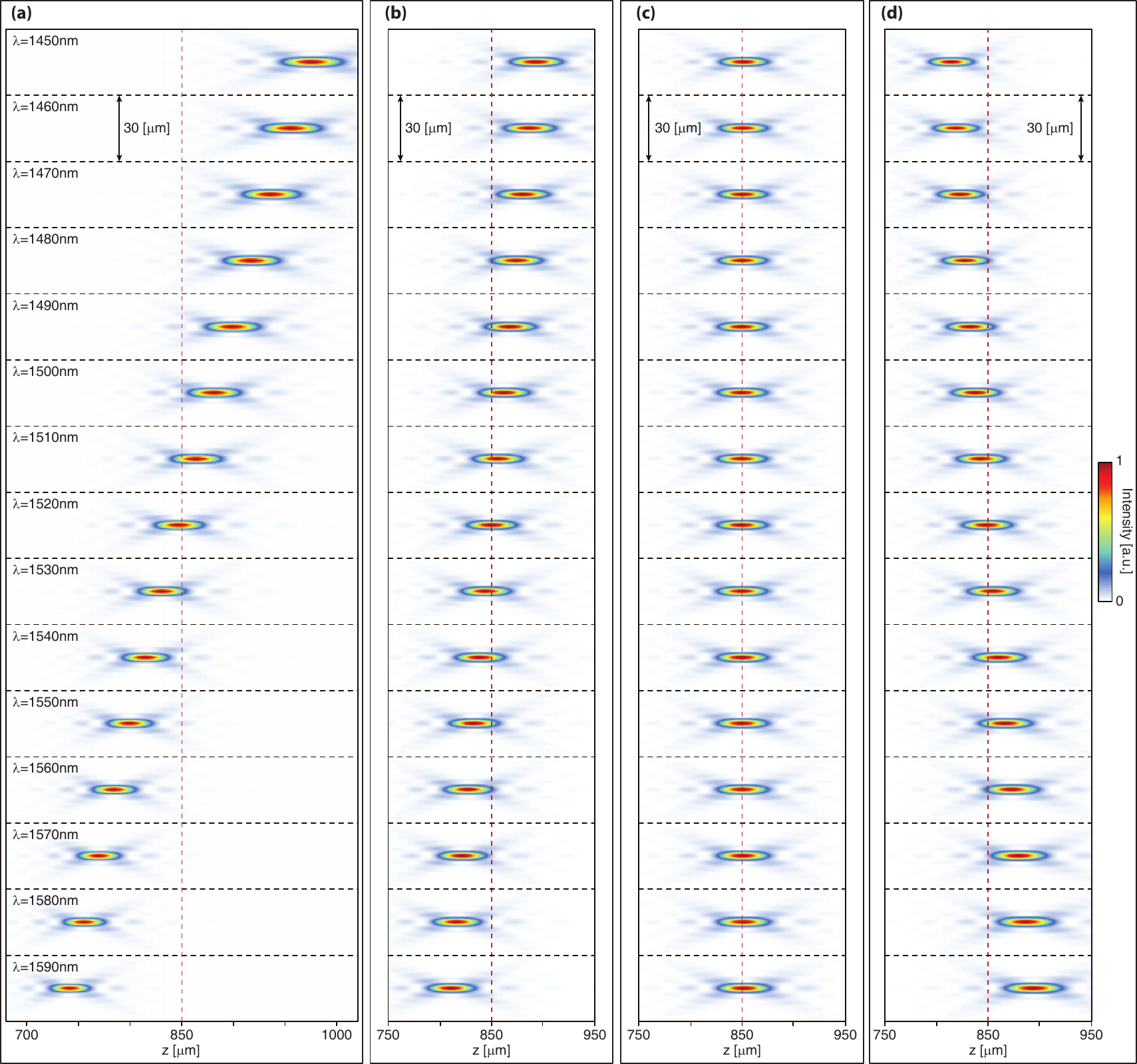}
\caption{Simulated axial intensity distribution for focusing mirrors with different dispersions designed using hypothetical meta-atoms. (a) Hyper-dispersive mirror. (b) Mirror with regular dispersion. (c) Mirror with zero dispersion. (d) Mirror with positive dispersion.}
\label{fig:S3_MadeDispersionLensAxailPlanes}
\end{figure*}

\clearpage

\begin{figure*}[htp]
\centering
\includegraphics[width=1\columnwidth]{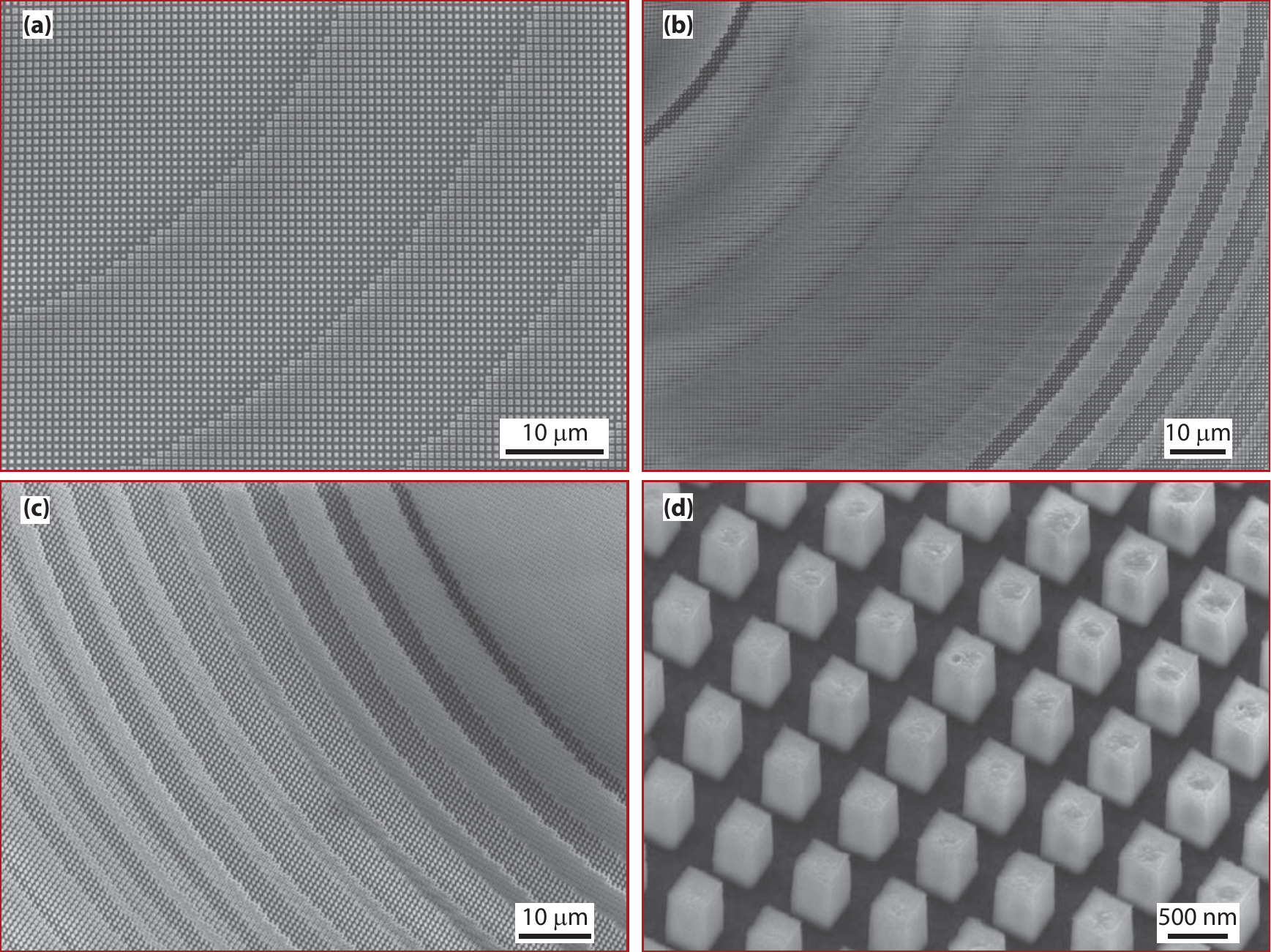}
\caption{Scanning electron micrographs of metasurface focusing mirrors with 850~$\mu$m focal distance. (a) Regular metasurface mirror. (b) Dispersionless metasurface mirror with $\sigma=300$~nm, and (c) $\sigma=50$~nm. (d) Fabricated meta-atoms.}
\label{fig:S4_SEMs}
\end{figure*}

\clearpage

\begin{figure*}[htp]
\centering
\includegraphics[width=1\columnwidth]{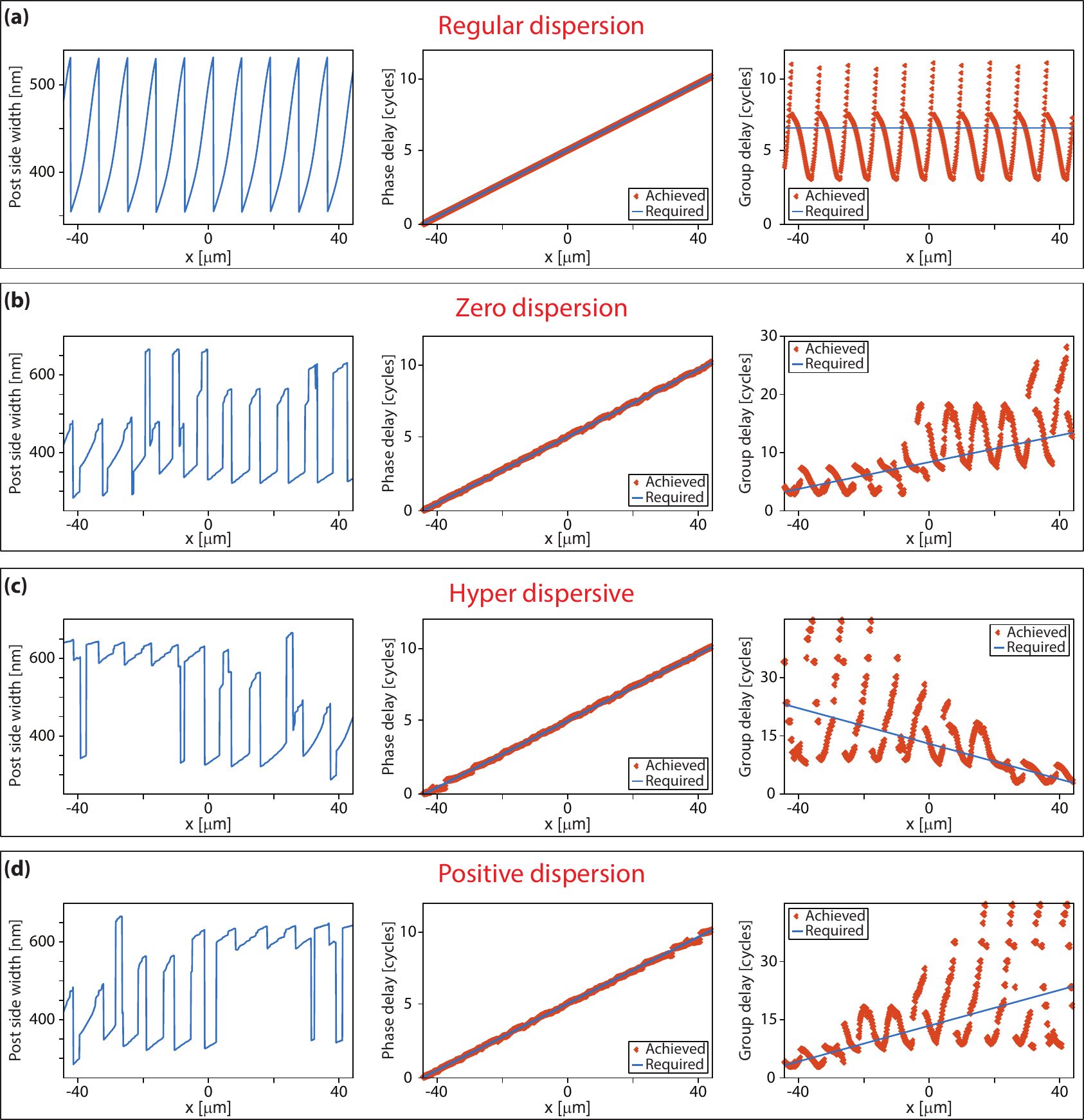}
\caption{Chosen nano-post side lengths for gratings and their corresponding phase and group delays. (a) The chosen nano-post side length (left), phase delay (center), and group delay (right) at 1520~nm for the fabricated regular grating. (b-d) Same as \textit{a} for the dispersionless, hyper-dispersive, and positive-dispersion gratings respectively.}
\label{fig:S5_GratingDelays}
\end{figure*}

\clearpage

\begin{figure*}[htp]
\centering
\includegraphics[width=1\columnwidth]{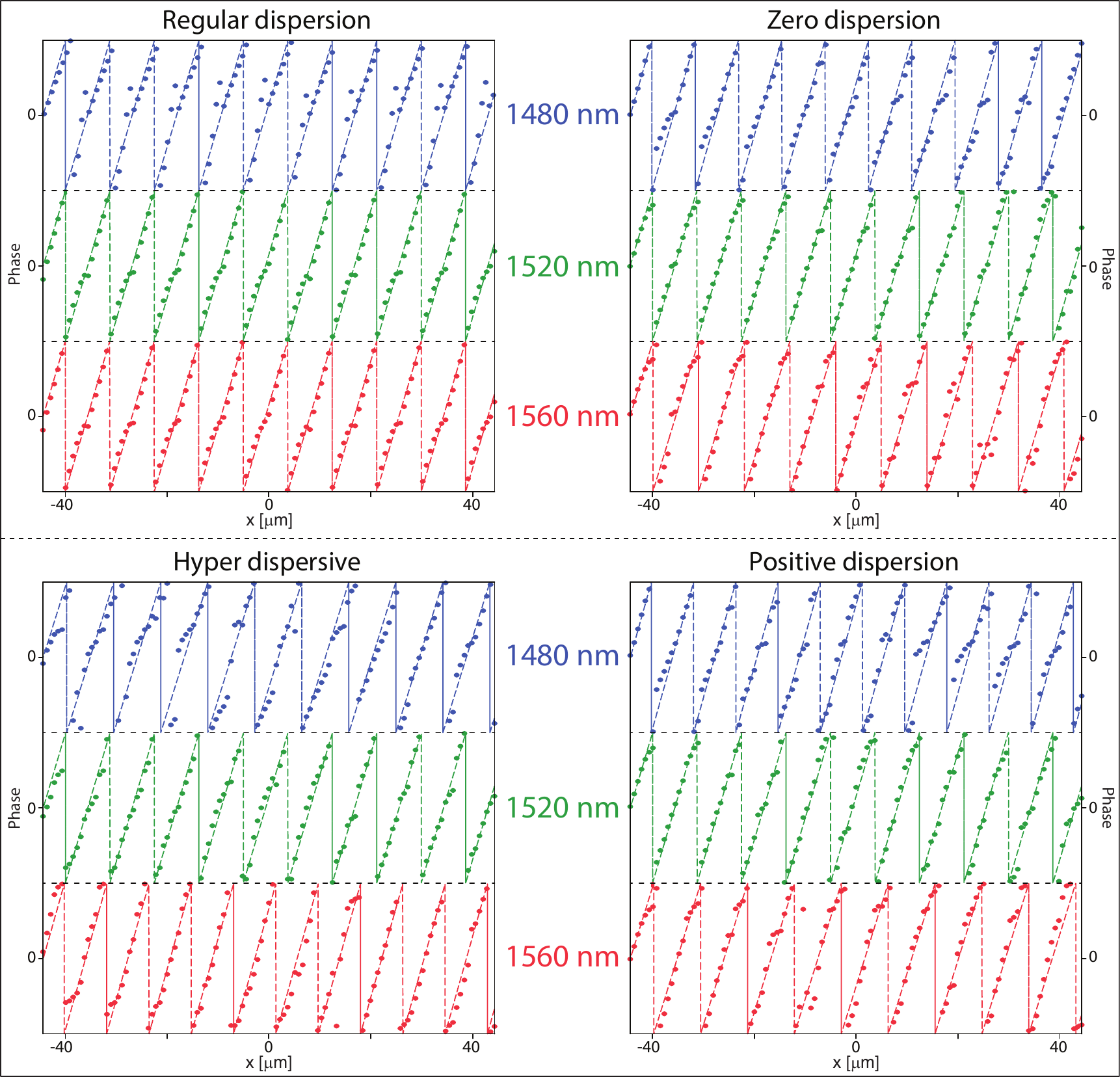}
\caption{Required and achieved phase values for gratings at three wavelengths. The phase delays are wrapped to the $-\pi$ to $\pi$ range. While the effective grating pitch is constant for the regular grating, it changes with wavelength in all other cases.}
\label{fig:S6_GratingPhasesThreeWavelengths}
\end{figure*}

\clearpage

\begin{figure*}[htp]
\centering
\includegraphics[width=1\columnwidth]{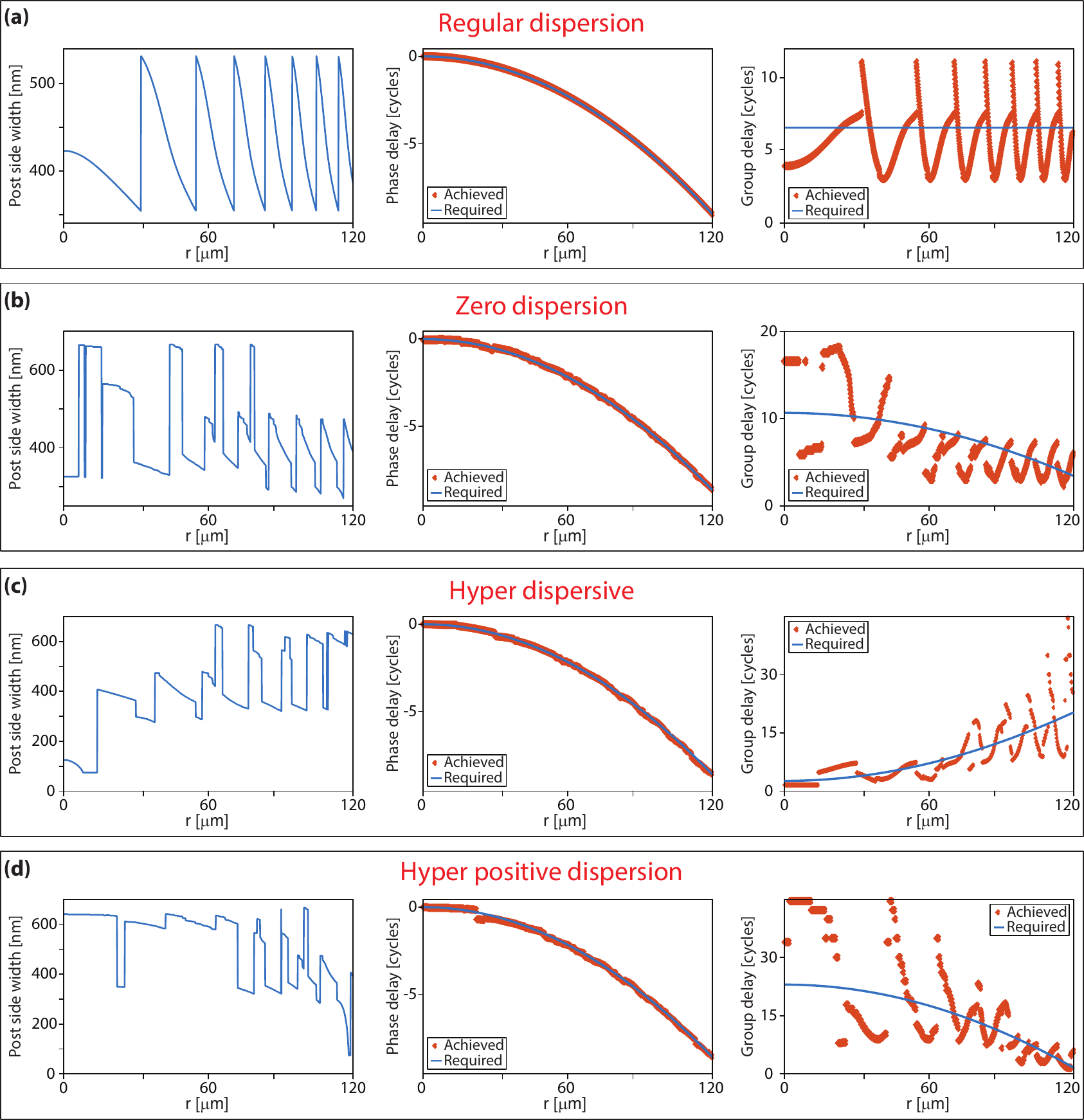}
\caption{Chosen nano-post side lengths for focusing mirrors and their corresponding phase and group delays. (a) The chosen nano-post side length (left), phase delay (center), and group delay (right) at 1520~nm for the fabricated 240~$\mu$m regular focusing mirror. (b-d) Same as \textit{a} for the dispersionless, hyper-dispersive, and positive-dispersion focusing mirrors respectively.}
\label{fig:S7_LensDelays}
\end{figure*}

\clearpage

\begin{figure*}[htp]
\centering
\includegraphics[width=1\columnwidth]{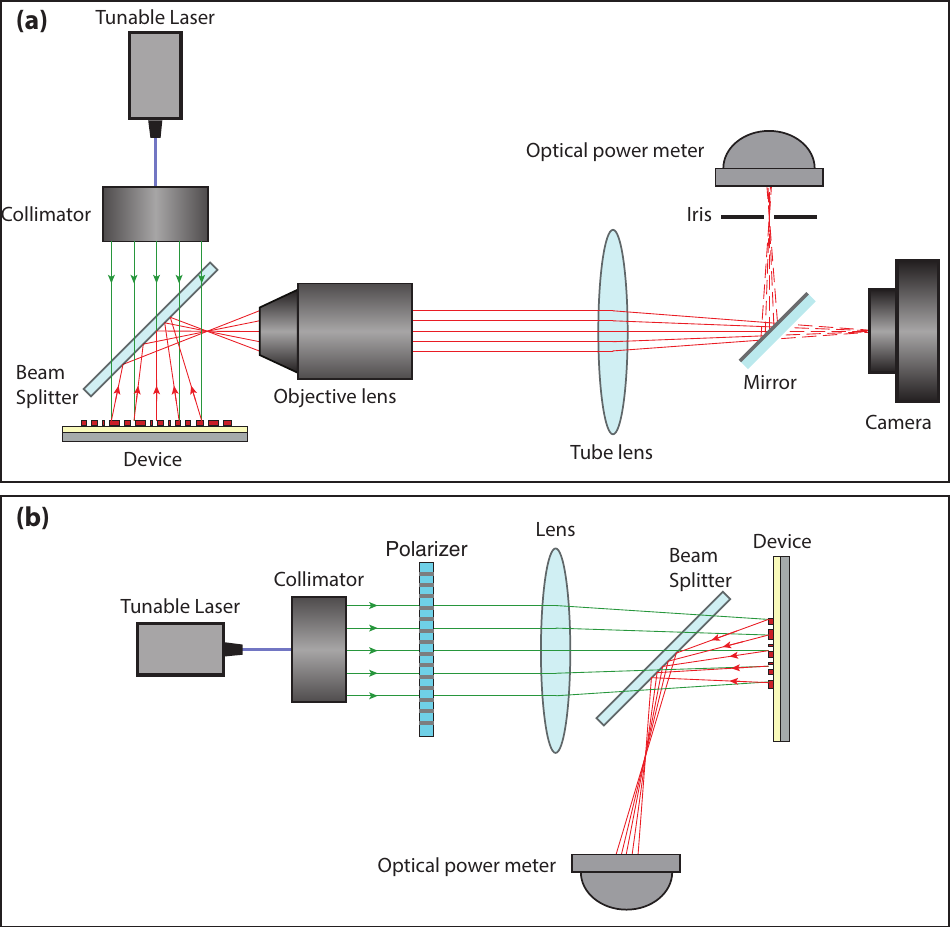}
\caption{Measurement setups. (a) Schematic illustration of the setup used to measure the deflection angles of gratings, and focus patterns and axial plane intensity distributions of focusing mirrors at different wavelengths. To measure the efficiency of the focusing mirrors, the flip mirror, iris, and optical power meter were used. (b) The setup used to measure the efficiencies of the gratings. The power meter was placed at a long enough distance such that the other diffraction orders fell safely outside its active aperture area.}
\label{fig:S8_MeasurementSetup}
\end{figure*}

\clearpage

\begin{figure*}[htp]
\centering
\includegraphics[width=1\columnwidth]{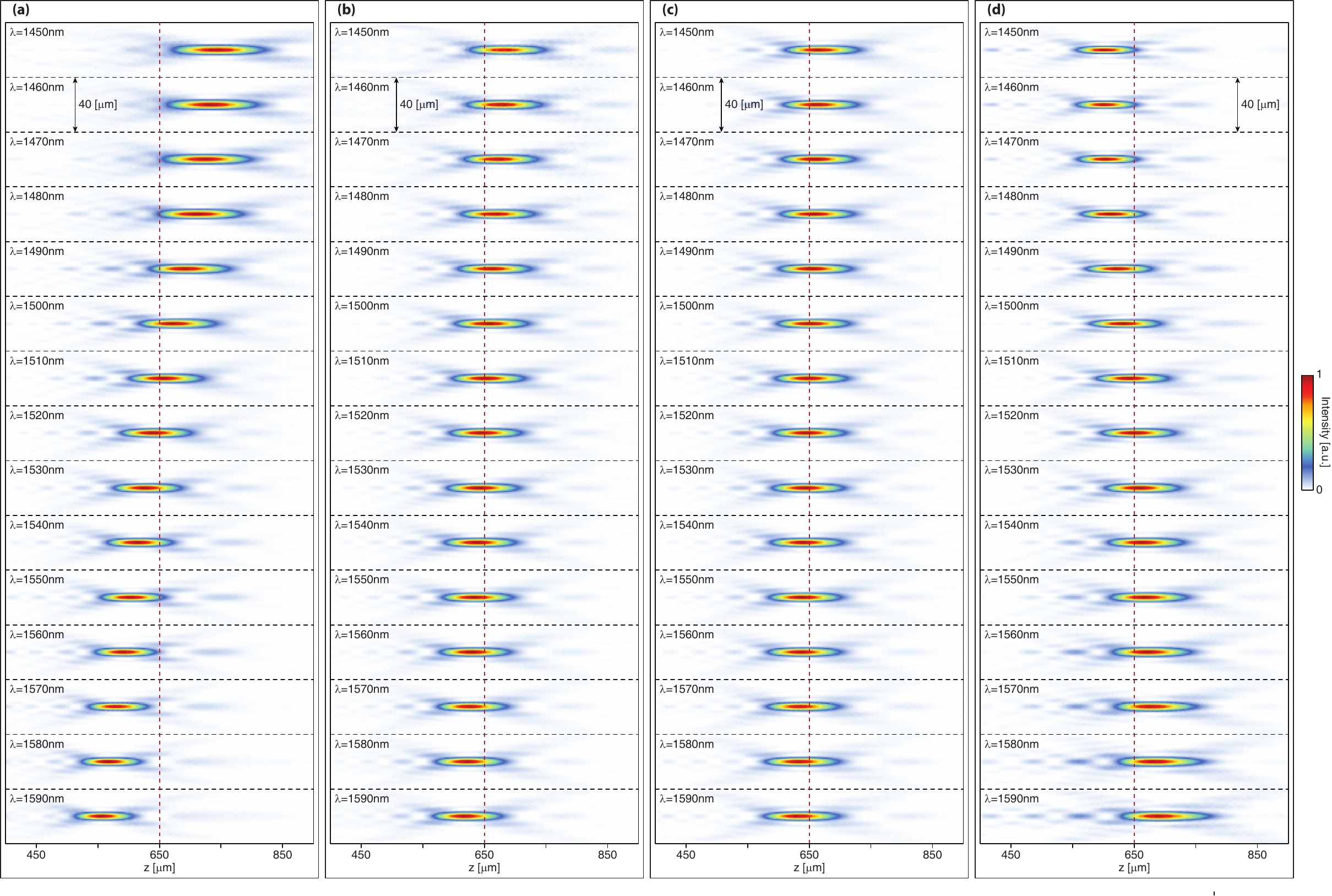}
\caption{Simulated axial intensity distribution for focusing mirrors with different dispersions designed using the reflective $\mathrm{\alpha}$-Si nano-posts discussed in Fig. 5(a). (a) Hyper-dispersive mirror. (b) Mirror with regular dispersion. (c) Mirror with zero dispersion. (d) Mirror with a positive dispersion with an amplitude twice the regular negative dispersion.}
\label{fig:S9_ActualPostsLensAxailPlanes_Sim}
\end{figure*}

\clearpage

\begin{figure*}[htp]
\centering
\includegraphics[width=1\columnwidth]{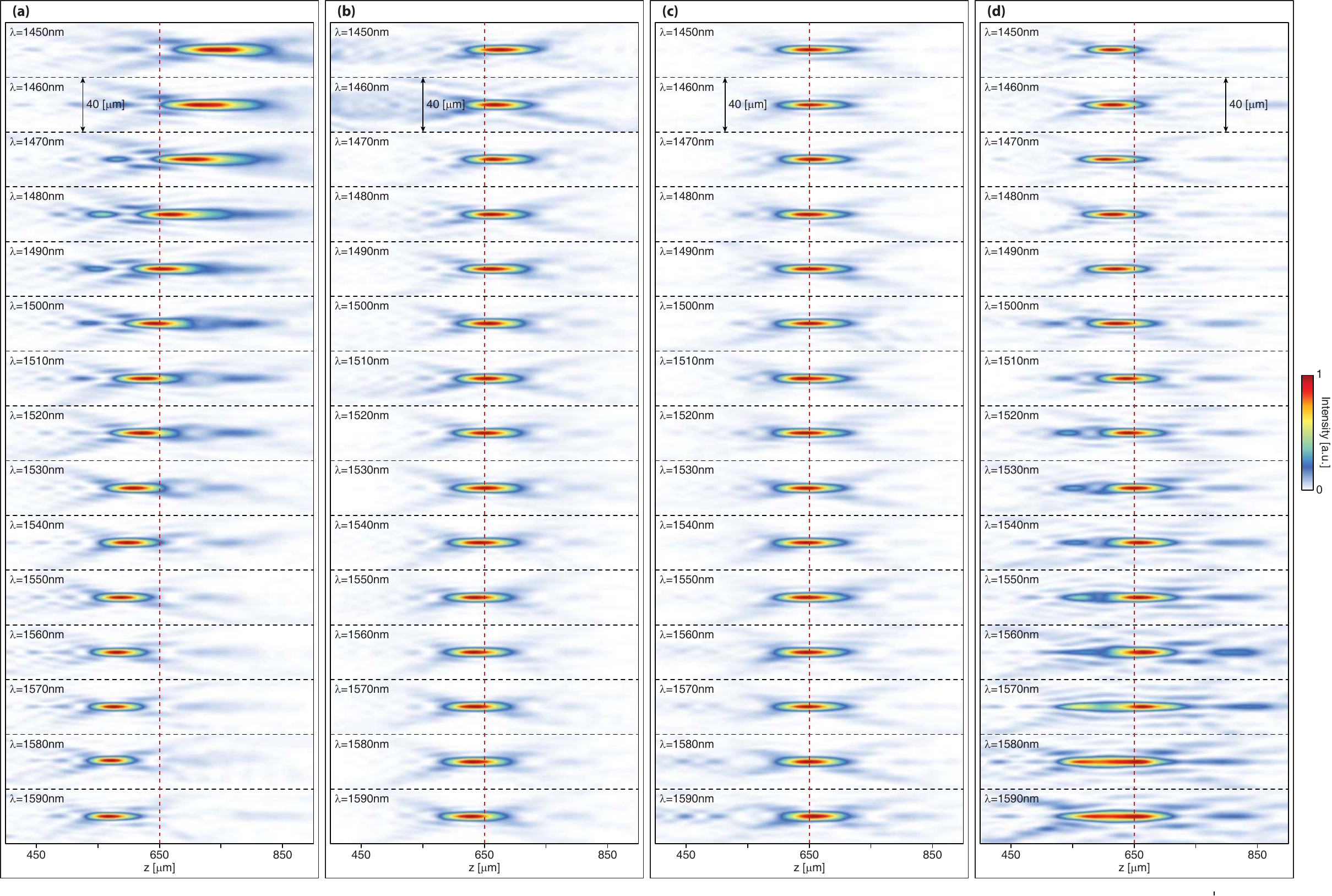}
\caption{Measured axial intensity distributions for focusing mirrors with different dispersions designed using the reflective $\mathrm{\alpha}$-Si nano-posts discussed in Fig. 5(b). (a), Hyper-dispersive mirror. (b) Mirror with regular dispersion. (c) Mirror with zero dispersion. (d) Mirror with a positive dispersion with an amplitude twice the regular negative dispersion.}
\label{fig:S10_ActualPostsLensAxailPlanes_Meas}
\end{figure*}

\clearpage

\begin{figure*}[htp]
\centering
\includegraphics[width=1\columnwidth]{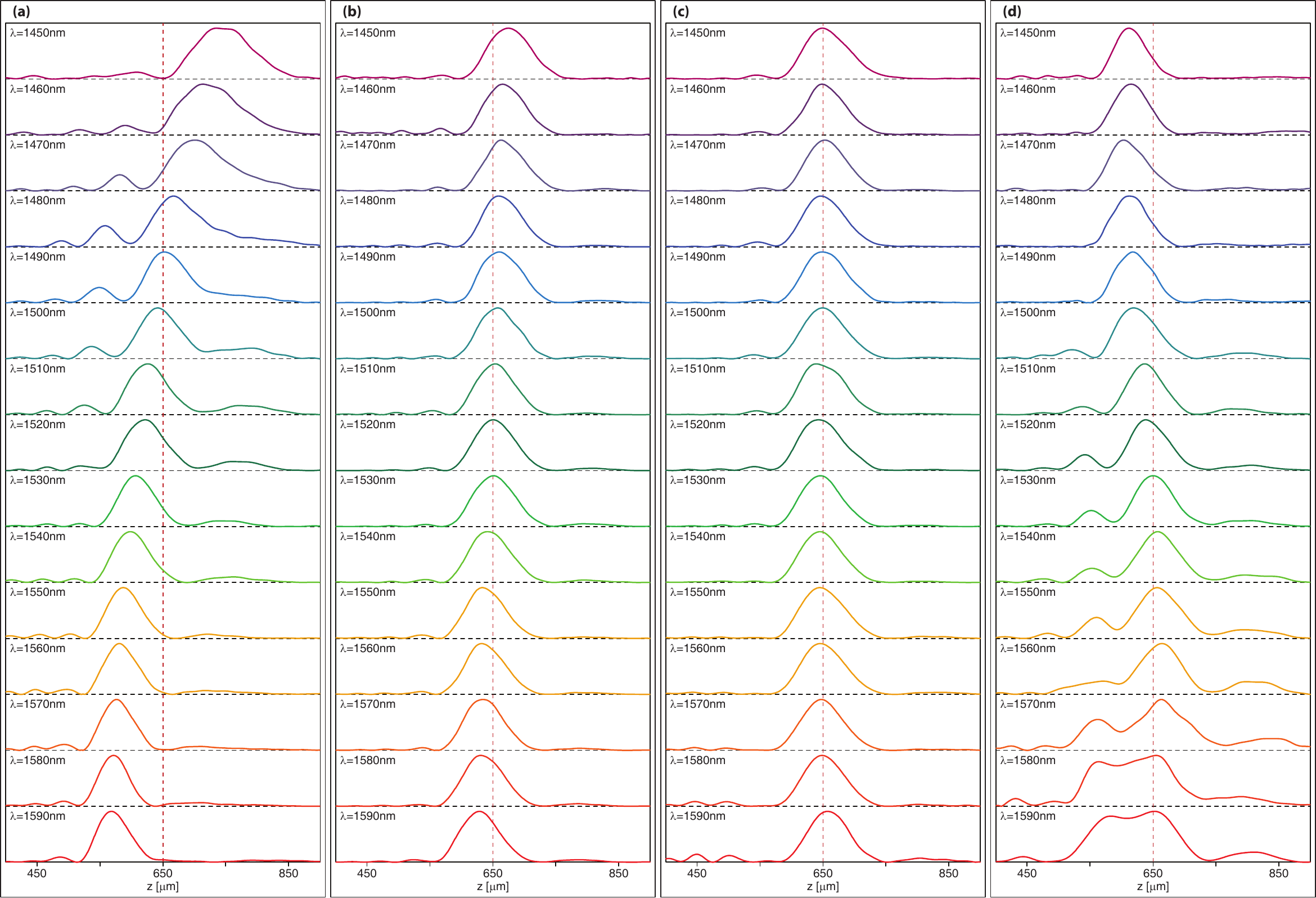}
\caption{One-dimensional cuts of the measured axial intensities plotted in Fig. \ref{fig:S10_ActualPostsLensAxailPlanes_Meas}. (a) Hyper-dispersive mirror. (b) Mirror with regular dispersion. (c) Mirror with zero dispersion. (d) Mirror with a positive dispersion with an amplitude twice the regular negative dispersion.}
\label{fig:S11_ActualPostsLensAxailPlanes_Meas_1D}
\end{figure*}

\clearpage

\begin{figure*}[htp]
\centering
\includegraphics[width=1\columnwidth]{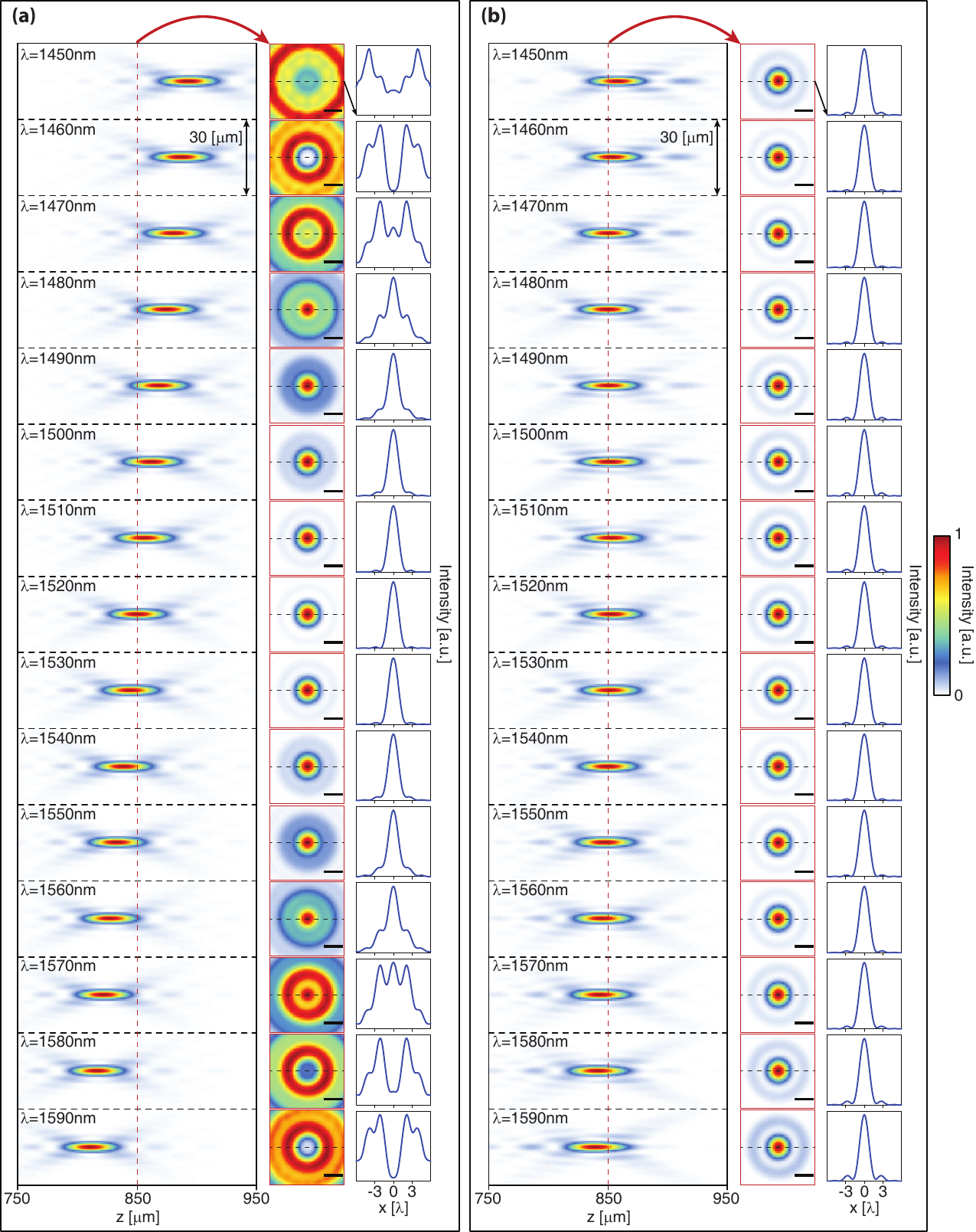}
\caption{Extended simulation results for the regular and dispersionless mirrors discussed in Figs. 5(c-g). (a) Simulated axial plane (left) and focal plane (center) intensities for a regular metasurface focusing mirror designed using the proposed reflective dielectric meta-atoms. One-dimensional cross-sections of the focal plane intensity is plotted on the right. The focusing mirror has a diameter of 500~$\mu$m and a focal distance of 850~$\mu$m at 1520~nm. (b) Similar results for a focusing mirror with the same parameters designed to have a minimal dispersion in the bandwidth. Scale bars: 2$\lambda$.}
\label{fig:S12_SupplementaryMeasurements}
\end{figure*}

\clearpage

\begin{figure*}[htp]
\centering
\includegraphics[width=1\columnwidth]{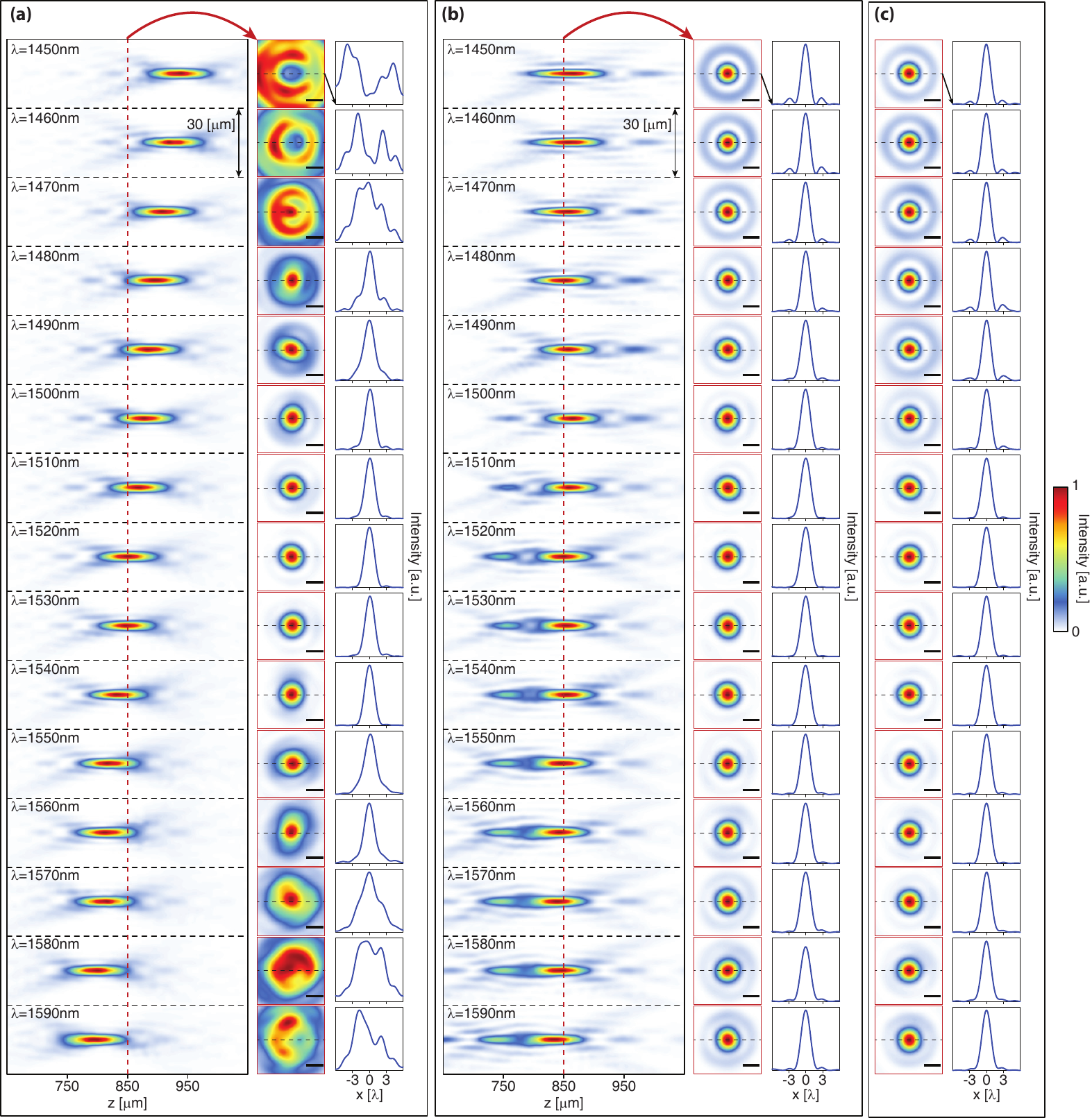}
\caption{Complete measurement results for the dispersionless and regular mirrors discussed in Figs. 5(c-g). (a) Measured intensities for the regular metasurface mirror. The axial plane intensities are shown on the left, the measured intensities in the 850~$\mu$m plane are plotted in the middle, and one dimensional cuts of the focal plane measurements are shown on the right. (b) Same as (a) but for the dispersionless mirror design with $\sigma=300$~nm. (c) Measured intensities in the plane 850~$\mu$m away from the surface of the dispersionless mirror with $\sigma=50$~nm. One dimensional cuts of the measured intensities are shown on the right. Scale bars: 2$\lambda$.}
\label{fig:S13_SupplementaryMeasurements}
\end{figure*}

\clearpage

\begin{figure*}[htp]
\centering
\includegraphics[width=1\columnwidth]{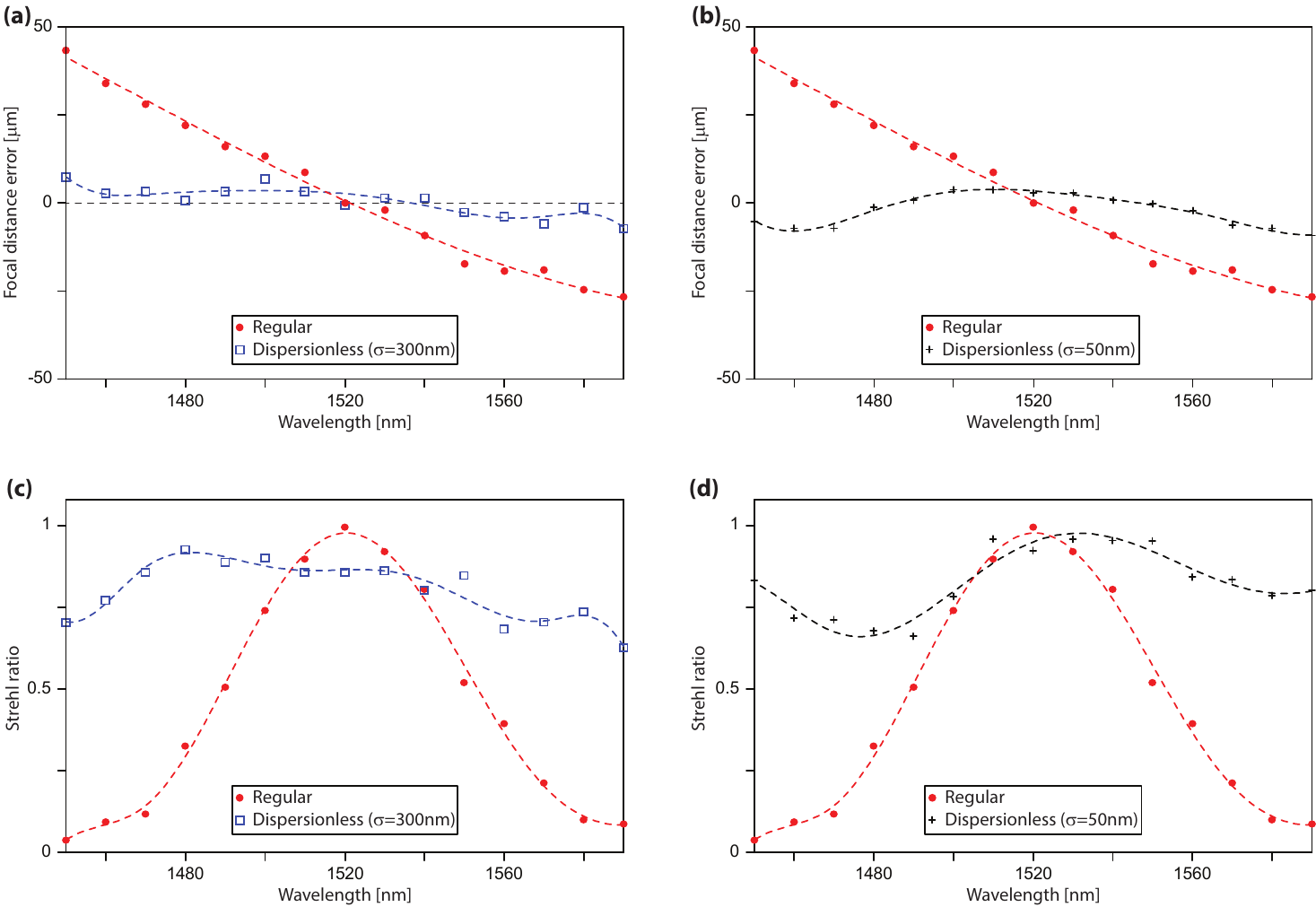}
\caption{Measured focal distances and Strehl ratios for the regular and dispersionless mirrors. (a) Measured focal distances for the regular and disperisonless ($\sigma =300$~nm) mirrors (same as Fig. 5(d)). (b) Measured focal distances for the regular and disperisonless ($\sigma =50$~nm) mirrors. (c) Strehl ratios calculated from the measured two dimensional modulation transfer functions (MTF) of the regular and dispersionless ($\sigma =300$~nm) metasurface mirrors. To find the Strehl ratio, the volume enclosed by the normalized two dimensional MTF is calculated at each wavelength. (d) The same graph as in (c), calculated and plotted for the $\sigma =50$~nm dispersionless mirror. In both cases, a clear flattening of the Strehl ratio, which is a measure of the contrast of an image formed by the mirror, is observed compared to the regular metasurface mirror.}
\label{fig:S14_StrehlRatios}
\end{figure*}

\clearpage

\begin{figure*}[ht]
\centering
\includegraphics[width=0.6\columnwidth]{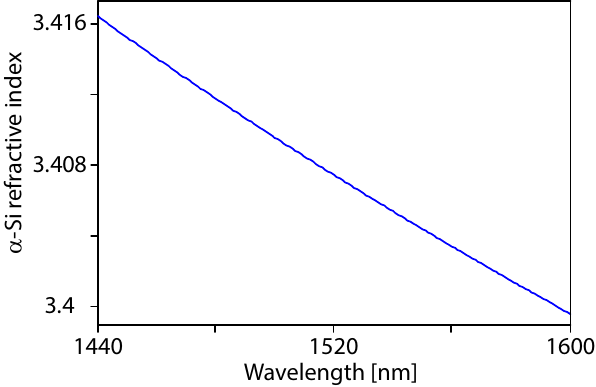}
\caption{Refractive index of amorphous silicon. The refractive index values were obtained using spectroscopic ellipsometry.}
\label{fig:S15_aSi_ind}
\end{figure*}


\begin{figure*}[ht]
\centering
\includegraphics[width=0.7\columnwidth]{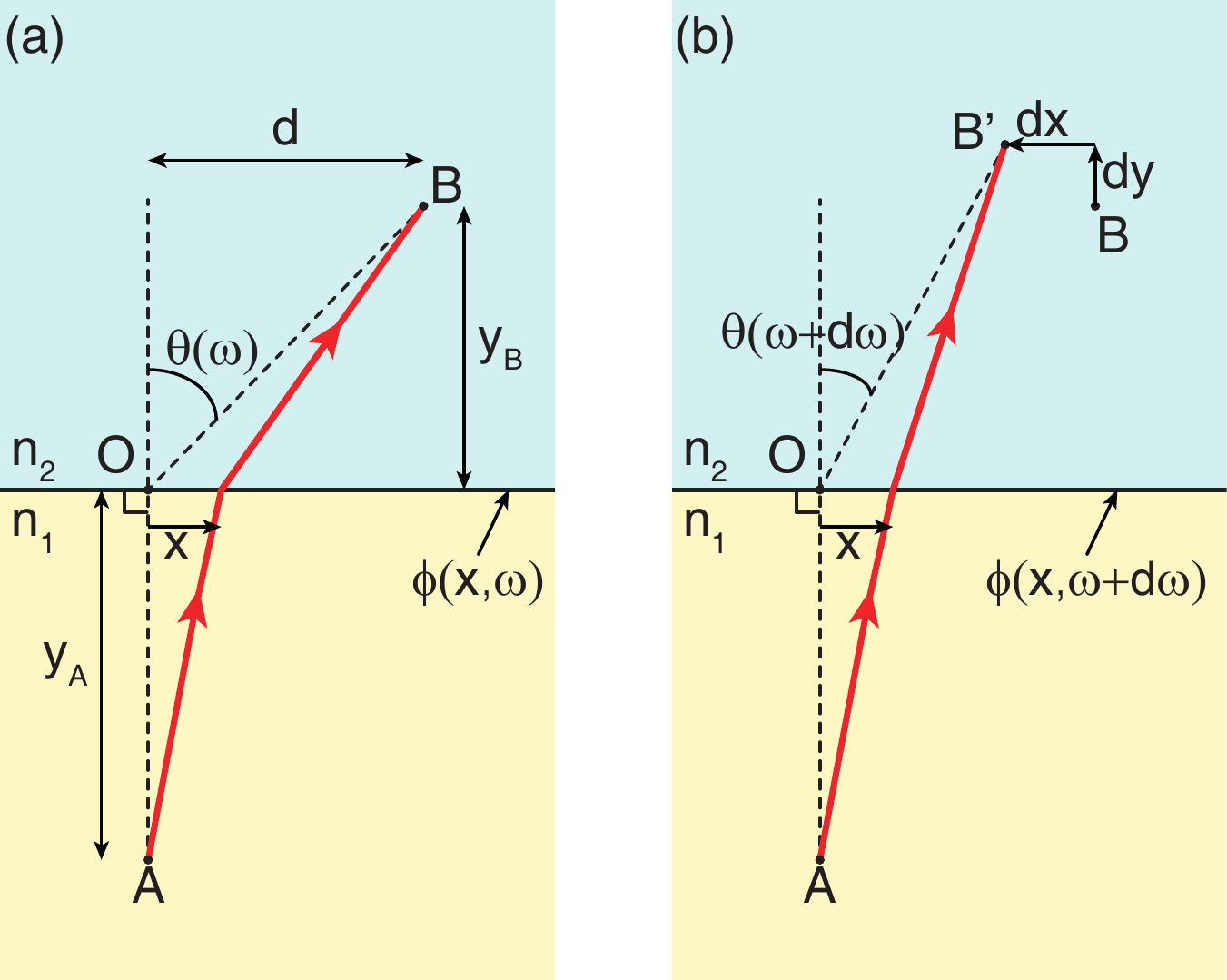}
\caption{Schematic of light deflection at a gradient phase surface. (a) A gradient phase surface between two materials with indices $\mathrm{n_1}$ and $\mathrm{n_2}$. At frequency $\omega$ a ray of light going from A to B, passes the interface at a point with coordinate $x$. (b) The same structure with a ray of light at $\omega+\mathrm{d}\omega$ that goes from A to B'.}
\label{fig:S16_Fermat}
\end{figure*}
\clearpage

\begin{figure*}[ht]
\centering
\includegraphics[width=0.5\columnwidth]{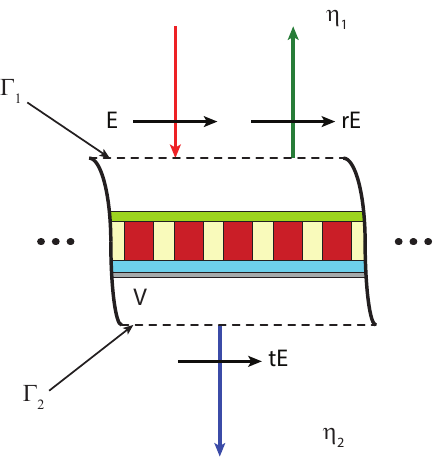}
\caption{Schematic of a generic metasurface. The metasurface is between two uniform materials with wave impedances of $\eta_1$ and $\eta_2$, and it is illuminated with a normally incident plane wave from the top side. Virtual planar boundaries $\Gamma_1$ and $\Gamma_2$ are used for calculating field integrals on each side of the metasurface.}
\label{fig:S17_PlanarStruct}
\end{figure*}
\clearpage





     }




\clearpage

\section*{Supplementary Information}

\section*{S1. Materials and Methods} 
\subsection*{Simulation and design.} The gratings with different dispersions discussed in Fig. 2(d) were designed using hypothetical meta-atoms that completely cover the required region of the phase-dispersion plane. We assumed that the meta-atoms provide 100 different phase steps from 0 to 2$\pi$, and that for each phase, 10 different dispersion values are possible, linearly spanning the 0 to $-150$~Rad$/\mu$m range. We assumed that all the meta-atoms have a transmission amplitude of 1. The design began with constructing the ideal phase masks at eight wavelengths equally spaced in the 1450 to 1590~nm range. This results in a vector of eight complex numbers for the ideal transmission at each point on the metasurface grating. The meta-atoms were assumed to form a two dimensional square lattice with a lattice constant of 740~nm, and one vector was generated for each lattice site. The optimum meta-atom for each site was then found by minimizing the Euclidean distance between the transmission vector of the meta-atoms and the ideal transmission vector for that site. The resulting phase mask of the grating was then found through a two-dimensional interpolation of the complex valued transmission coefficients of the chosen meta-atoms. The grating area was assumed to be illuminated uniformly, and the deflection angle of the grating was found by taking the Fourier transform of the field after passing through the phase mask, and finding the angle with maximum intensity. A similar method was used to design and simulate the focusing mirrors discussed in Figs. 2(e-i). In this case, the meta-atoms are assumed to cover dispersion values up to $-200$~Rad$/\mu$m. The meta-atoms provide 21 different dispersion values distributed uniformly in the 0 to $-200$~Rad$/\mu$m range. The focusing mirrors were designed and the corresponding phase masks were found in a similar manner to the gratings. A uniform illumination was used as the source, and the resulting field after reflection from the mirror was propagated in free space using a plane wave expansion method to find the intensity in the axial plane. The focal distances plotted in Fig. 2(e) show the distance of the maximum intensity point from the mirrors at each wavelength. The gratings and focusing mirrors discussed in Figs. 4(a), 5(a), and 5(c) are designed and simulated in exactly the same manner, except for using actual dielectric meta-atom reflection amplitudes and phases instead of the hypothetical ones.

If the actual meta-atoms provided an exactly linear dispersion (i.e. if their phase was exactly linear with frequency over the operation bandwidth), one could use the required values of the phase and dispersion at each lattice site to choose the best meta-atom (knowing the coordinates of one point on a line and its slope would suffice to determine the line exactly). The phases of the actual meta-atoms, however, do not follow an exactly linear curve [Fig. 3(d)]. Therefore, to minimize the error between the required phases, and the actual ones provided by the meta-atoms, we have used a minimum weighted Euclidean distance method to design the devices fabricated and tested in the manuscript: at each point on the metasurface, we calculate the required complex reflection at eight wavelengths (1450~nm to 1590~nm, at 20~nm distances). We also calculate the complex reflection provided by each nano-post at the same wavelengths. To find the best meta-atom for each position, we calculate the weighted Euclidean distance between the required reflection vector, and the reflection vectors provided by the actual nano-posts. The nano-post with the minimum distance is chosen at each point. As a result, the chromatic dispersion is indirectly taken into account, not directly. The weight function can be used to increase or decrease the importance of each part of the spectrum depending on the specific application. In this work, we have chosen an inverted Gaussian weight function ($\exp((\lambda-\lambda_0)^2/2\sigma^2)$, $\lambda_0=1520$~nm, $\sigma=300$~nm) for all the devices to slightly emphasize the importance of wavelengths farther from the center. In addition, we have also designed a dispersionless lens with $\sigma=50$~nm (the measurement results of which are provided in Figs. \ref{fig:S13_SupplementaryMeasurements} and \ref{fig:S14_StrehlRatios}) for comparison. The choice of 8 wavelengths to form and compare the reflection vectors is relatively arbitrary; however, the phases of the nano-posts versus wavelength are smooth enough, such that they can be well approximated by line segments in 20~nm intervals. In addition, performing the simulations at 8 wavelengths is computationally not very expensive. Therefore, 8 wavelengths are enough for a 150~nm bandwidth here, and increasing this number may not result in a considerable improvement in the performance.

Reflection amplitude and phase of the meta-atoms were found using rigorous coupled wave analysis technique~\cite{Liu2012CompPhys}. For each meta-atom size, a uniform array on a subwavelength lattice was simulated using a normally incident plane wave. The subwavelength lattice ensures the existence of only one propagating mode which justifies the use of only one amplitude and phase for describing the optical behavior at each wavelength. In the simulations, the amorphous silicon layer was assumed to be 725~nm thick, the SiO$_{2}$ layer was 325~nm, and the aluminum layer was 100~nm thick. A 30-nm-thick Al$_{2}$O$_{3}$ layer was added between the Al and the oxide layer (this layer served as an etch stop layer to avoid exposing the aluminum layer during the etch process). Refractive indices were set as follows in the simulations: SiO$_{2}$: 1.444, Al$_{2}$O$_{3}$: 1.6217, and Al: 1.3139-$i$13.858. The refractive index of amorphous silicon used in the simulations is plotted in Fig. \ref{fig:S15_aSi_ind}.

The FDTD simulations of the gratings (Figs. 4(e-h)) were performed using a normally incident plane-wave illumination with a Gaussian amplitude in time (and thus a Gaussian spectrum) in MEEP~\cite{Oskooi2010CompPhys}. The reflected electric field was saved in a plane placed one wavelength above the input plane at time steps of 0.05 of the temporal period. The results in Figs. 4(e-h) are obtained via Fourier transforming the fields in time and space resulting in the reflection intensities as a function of frequency and transverse wave-vector.

\subsection*{Sample fabrication.} A 100-nm aluminum layer and a 30-nm Al$_{2}$O$_{3}$ layer were deposited on a silicon wafer using electron beam evaporation. This was followed by deposition of 325~nm of SiO$_{2}$ and 725~nm of amorphous silicon using the plasma enhanced chemical vapor deposition (PECVD) technique at $200\,^{\circ}{\rm C}$. A $\sim$300~nm thick layer of ZEP-520A positive electron-beam resist was spun on the sample at 5000 rpm for 1 min, and was baked at $180\,^{\circ}{\rm C}$ for 3 min. The pattern was generated using a Vistec EBPG5000+ electron beam lithography system, and was developed for 3 minutes in the ZED-N50 developer (from Zeon Chemicals). A $\sim$70-nm Al$_{2}$O$_{3}$ layer was subsequently evaporated on the sample, and the pattern was reversed with a lift off process. The Al$_{2}$O$_{3}$ hard mask was then used to etch the amorphous silicon layer in a 3:1 mixture of $\mathrm{SF_6}$ and $\mathrm{C_4F_8}$ plasma. The mask was later removed using a 1:1 solution of ammonium hydroxide and hydrogen peroxide at $80^{\circ}$~C.

\subsection*{Measurement procedure.} The measurement setup is shown in Fig. \ref{fig:S8_MeasurementSetup}(a). Light emitted from a tunable laser source (Photonetics TUNICS-Plus) was collimated using a fiber collimation package (Thorlabs F240APC-1550), passed through a 50/50 beamsplitter (Thorlabs BSW06), and illuminated the device. For grating measurements a lens with a 50~mm focal distance was also placed before the grating at a distance of $\sim$45~mm to partially focus the beam and reduce the beam divergence after being deflected by the grating in order to decrease the measurement error (similar to Fig. \ref{fig:S8_MeasurementSetup}(b)). The light reflected from the device was redirected using the same beamsplitter, and imaged using a custom built microscope. The microscope consists of a 50X objective (Olympus LMPlanFL N, NA=0.5), a tube lens with a 20 cm focal distance (Thorlabs AC254-200-C-ML), and an InGaAs camera (Sensors Unlimited 320HX-1.7RT). The grating deflection angle was found by calculating the center of mass for the deflected beam imaged 3~mm away from the gratings surface. For efficiency measurements of the focusing mirrors, a flip mirror was used to send light towards an iris (2 mm diameter, corresponding to an approximately 40~$\mu$m iris in the object plane) and a photodetector (Thorlabs PM100D with a Thorlabs S122C head). The efficiencies were normalized to the efficiency of the regular mirror at its center wavelength by dividing the detected power through the iris by the power measured for the regular mirror at its center wavelength. The measured intensities were up-sampled using their Fourier transforms in order to achieve smooth intensity profiles in the focal and axial planes. To measure the grating efficiencies, the setup shown in Supporting Information Fig. \ref{fig:S8_MeasurementSetup}(b) was used, and the photodetector was placed $\sim$50~mm away from the grating, such that the other diffraction orders fall outside its active area. The efficiency was found by calculating the ratio of the power deflected by the grating to the power normally reflected by the aluminum reflector in areas of the sample with no grating. The beam-diameter on the grating was calculated using the setup parameters, and it was found that $\sim$84$\%$ of the power was incident on the 90~$\mu$m wide gratings. This number was used to correct for the lost power due to the larger size of the beam compared to the grating.

\section*{S2. Chromatic dispersion of diffractive devices.} 
Chromatic dispersion of a regular diffractive grating or lens is set by its function. The grating momentum for a given order of a grating with a certain period is constant and does not change with changing the wavelength. If we denote the size of the grating reciprocal lattice vector of interest by $k_G$, we get:
\begin{equation}
\sin(\theta) = \frac{k_G}{2\pi/\lambda} \Rightarrow \theta=\sin^{-1}(\frac{k_G}{2\pi/\lambda}),
\label{eq:GratingDeflectionAngle}\\
\end{equation}
where $\theta$ is the deflection angle at a wavelength $\lambda$ for normally incident beam. The chromatic angular dispersion of the grating ( ${\mathrm{d}\theta}/{\mathrm{d}\lambda}$) is then given by:
\begin{equation}
\frac{\mathrm{d}\theta}{\mathrm{d}\lambda} = \frac{k_G/2\pi}{\sqrt{1-(k_G\lambda/2\pi)^2}}=\frac{\tan(\theta)}{\lambda}.
\label{eq:GratingDispersion}\\
\end{equation}
\noindent{and in terms of frequency:}
\begin{equation}
\frac{\mathrm{d}\theta}{\mathrm{d}\omega} = -\frac{\tan(\theta)}{\omega}.
\label{eq:GratingDispersion_frequency}\\
\end{equation}
Therefore, the dispersion of a regular grating only depends on its deflection angle and the wavelength. Similarly, focal distance of one of the focal points of diffractive and metasurface lenses changes as ${\mathrm{d}f}/{\mathrm{d}\lambda}=-f/\lambda$ (thus ${\mathrm{d}f}/{\mathrm{d}\omega}=f/\omega$ (\cite{Born1999,Arbabi2016Optica,Faklis1995ApplOpt}).

\section*{S3. Chromatic dispersion of multiwavelength diffractive devices.} 
As it is mentioned in the main text, multiwavelength diffractive devices~(\cite{Arbabi2016Optica,Faklis1995ApplOpt,Aieta2015Science}) do not change the dispersion of a given order in a grating or lens. They are essentially multi-order gratings or lenses, where each order has the regular (negative) diffractive chromatic dispersion. These devices are designed such that at certain distinct wavelengths of interest, one of the orders has the desired deflection angle or focal distance. If the blazing of each order at the corresponding wavelength is perfect, all of the power can be directed towards that order at that wavelength. However, at wavelengths in between the designed wavelengths, where the grating or lens is not corrected, the multiple orders have comparable powers, and show the regular diffractive dispersion. This is schematically shown in Fig. \ref{fig:S1_Multi_vs_Ach}(a). Figure \ref{fig:S1_Multi_vs_Ach}(b) compares the chromatic dispersion of a multi-wavelength diffractive lens to a typical refractive apochromatic lens.

\section*{S4. Generalization of chromatic dispersion control to nonzero dispersions.} 
Here we present the general form of equations for the dispersion engineered metasurface diffractive devices. We assume that the function of the device is set by a parameter $\xi (\omega)$, where we have explicitly shown its frequency dependence. For instance, $\xi$ might denote the deflection angle of a grating or the focal distance of a lens. The phase profile of a device with a desired $\xi (\omega)$ is given by
\begin{equation}
\phi(x,y,\xi(\omega);\omega) = \omega T(x,y,\xi(\omega)),
\label{eq:arb_disp_phase}\\
\end{equation}
which is the generalized form of the Eq. (1). We are interested in controlling the parameter $\xi (\omega)$ and its dispersion (i.e. derivative) at a given frequency $\omega_0$.  $\xi (\omega)$ can be approximated as $\xi(\omega)\approx\xi_0+\partial\xi/\partial\omega|_{\omega=\omega_0}(\omega-\omega_0)$ over a narrow bandwidth around $\omega_0$. Using this approximation, we can rewrite \ref{eq:arb_disp_phase} as
\begin{equation}
\phi(x,y;\omega) = \omega T(x,y,\xi_0+\partial\xi/\partial\omega|_{\omega=\omega_0}(\omega-\omega_0)).
\label{eq:linear_disp_phase}\\
\end{equation}

At $\omega_0$, this reduces to
\begin{equation}
\phi(x,y;\omega)|_{\omega=\omega_0} = \omega_0 T(x,y,\xi_0),
\label{eq:phase_disp_phase}\\
\end{equation}
and the phase dispersion at $\omega_0$ is given by
\begin{equation}
\frac{\mathrm{\partial}\phi(x,y;\omega)}{\mathrm{\partial}\omega}|_{\omega=\omega_0} = T(x,y,\xi_0)+\partial\xi/\partial\omega|_{\omega=\omega_0}\omega_0\frac{\partial T(x,y,\xi)}{\partial\xi}|_{\xi=\xi_0}.
\label{eq:disp_disp_phase}\\
\end{equation}

Based on Eqs. (\ref{eq:phase_disp_phase}) and (\ref{eq:disp_disp_phase}) the values of $\xi_0$ and $\partial\xi/\partial\omega|_{\omega=\omega_0}$ can be set independently, if the phase $\phi(x,y,\omega_0)$ and its derivative $\partial\phi/\partial\omega$ can be controlled simultaneously and independently. Therefore, the device function at $\omega_0$ (determined by the value of $\xi_0$)  and its dispersion (determined by $\partial\xi/\partial\omega|_{\omega=\omega_0}$) will be decoupled. The zero dispersion case is a special case of Eq. (\ref{eq:disp_disp_phase}) with $\partial\xi/\partial\omega|_{\omega=\omega_0}=0$. In the following we apply these results to the special cases of blazed gratings and spherical-aberration-free lenses (also correct for spherical-aberration-free focusing mirrors).

For a 1-dimensional conventional blazed grating we have $\xi=\theta$ (the deflection angle), and $T =- x\sin(\theta)$. Therefore the phase profile with a general dispersion is given by:
\begin{equation}
\phi(x;\omega) = -\omega x \sin[\theta_0+ D(\omega-\omega_0)],
\label{eq:Generalized_grating_phase_linear_freq}\\
\end{equation}
\noindent where $D = \partial\theta/\partial\omega|_{\omega=\omega_0} = \nu D_0$, and $D_0=-\tan(\theta_0)/\omega_0$ is the angular dispersion of a regular grating with deflection angle $\theta_0$ at the frequency $\omega_0$. We have chosen to express the generalized dispersion $D$ as a multiple of the regular dispersion $D_0$ with a real number $\nu$ to benchmark the change in dispersion. For instance, $\nu=1$ corresponds to a regular grating, $\nu=0$ represents a dispersionless grating, $\nu=-1$ denotes a grating with positive dispersion, and $\nu=3$ results in a grating three times more dispersive than a regular grating (i.e. hyper-dispersive). Various values of $\nu$ can be achieved using the method of simultaneous control of phase and dispersion of the meta-atoms, and thus we can break this fundamental relation between the deflection angle and angular dispersion. The phase derivative necessary to achieve a certain value of $\nu$ is given by:
\begin{equation}
\frac{\partial\phi(x;\omega)}{\partial\omega}|_{\omega=\omega_0} = -x/c \sin(\theta_0)(1-\nu),
\label{eq:Generalized_grating_disp_linear_freq}\\
\end{equation}
\noindent or in terms of wavelength:
\begin{equation}
\frac{\partial\phi(x;\lambda)}{\partial\lambda}|_{\lambda=\lambda_0} = \frac{2\pi}{{\lambda_0}^2} x \sin(\theta_0)(1-\nu).
\label{eq:Generalized_grating_disp_linear}\\
\end{equation}
For a spherical-aberration-free lens we have $\xi=f$ and $T(x,y,f)=-\sqrt{x^2 +y^2 + f^2}/c$. Again we can approximate $f$ with its linear approximation $f(\omega)=f_0+D(\omega-\omega_0)$, with $D=\partial f/ \partial \omega|_{\omega=\omega_0}$ denoting the focal distance dispersion at $\omega=\omega_0$. The regular dispersion for such a lens is given by $D_0=f_0/\omega_0$. Similar to the gratings, we can write the more general form for the focal distance dispersion as $D=\nu D_0$, where $\nu$ is some real number. In this case, the required phase dispersion is given by:
\begin{equation}
\frac{\partial\phi(x,y;\omega)}{\partial\omega}|_{\omega=\omega_0} = -\frac{1}{c}[\sqrt{x^2 +y^2 + {f_0}^2}+\frac{\nu {f_0}^2}{\sqrt{x^2 +y^2 + {f_0}^2}}],
\label{eq:Generalized_lens_disp_linear_freq}\\
\end{equation}
\noindent which can also be expressed in terms of wavelength:
\begin{equation}
\frac{\partial\phi(x,y;\lambda)}{\partial\lambda}|_{\lambda=\lambda_0} = \frac{2\pi}{{\lambda_0}^2}[\sqrt{x^2 +y^2 + {f_0}^2}+\frac{\nu {f_0}^2}{\sqrt{x^2 +y^2 + {f_0}^2}}].
\label{eq:Generalized_lens_disp_linear}\\
\end{equation}

\section*{S5. Maximum meta-atom dispersion required for controlling chromatic dispersion of gratings and lenses.} 
Since the maximum achievable dispersion is limited by the meta-atom design, it is important to find a relation between the maximum dispersion required for implementation of a certain metasurface device, and the device parameters (e.g. size, focal distance, deflection angle, etc.). Here we find these maxima for the cases of gratings and lenses with given desired dispersions.

For the grating case, it results from Eq. (\ref{eq:Generalized_grating_disp_linear}) that the maximum required dispersion is given by
\begin{equation}
\mathrm{max}(\frac{\partial\phi(x;\lambda)}{\partial\lambda}|_{\lambda=\lambda_0}) = k_0 X \frac{\sin(\theta_0)}{\lambda_0}(1-\nu),
\label{eq:Grating_max_dispersion}\\
\end{equation}
where $X$ is the length of the grating, and $k_0=2 \pi/\lambda_0$ is the wavenumber. It is important to note that based on the value of $\nu$, the sign of the meta-atom dispersion changes. However, in order to ensure a positive group velocity for the meta-atoms, the dispersions should be negative. Thus, if $1-\nu>0$, a term should be added to make the dispersion values negative. We can always add a term of type $\phi_0=k L_0$ to the phase without changing the function of the device. This term can be used to shift the required region in the phase-dispersion plane. Therefore, it is actually the difference between the minimum and maximum of Eqs. \ref{eq:Generalized_grating_disp_linear} and \ref{eq:Generalized_lens_disp_linear} that sets the maximum required dispersion. Using a similar procedure, we find the maximum necessary dispersion for a spherical-aberration-free lens as
\begin{equation}
\phi'_{\mathrm{max}} = -\frac{k_0 f}{\lambda_0}\begin{cases}

\frac{\Theta+\nu}{\sqrt{\Theta}}-1-\nu & \nu<1 \\

\frac{\Theta+\nu}{\sqrt{\Theta}}-2\sqrt{\nu}  & 1<\nu<\sqrt{\Theta}\\

(1-\sqrt{\nu})^2  & \sqrt{\Theta}<\nu<\Theta\\

-(\frac{\Theta+\nu}{\sqrt{\Theta}}-1-\nu) &\Theta<\nu\\

\end{cases},
\label{eq:Lens_max_disp}
\end{equation}
where $f$ is the focal distance of the lens, and $\Theta=(f^2+R^2)/f^2=1/(1-\mathrm{NA}^2)$ ( R: lens radius, NA: numerical aperture). $\log{[\phi'_{\mathrm{max}}/( -k_0 f/\lambda_0)]}$ is plotted in Fig. \ref{fig:S2_MaxDisps}(a) as a function of NA and $\nu$. In the simpler case of dispersionless lenses (i.e. $\nu=0$), Eq. (\ref{eq:Lens_max_disp}) can be further simplified to
\begin{equation}
\phi'_{\mathrm{max}} =-\frac{k_0 R}{\lambda}\frac{1-\sqrt{1-\mathrm{NA}^2}}{\mathrm{NA}}\approx -\frac{k_0 R \mathrm{NA}}{2\lambda}
\label{eq:max_dispersion_simplified}\\
\end{equation}
where $R$ is the lens radius and the approximation is valid for small values of NA. The maximum required dispersion for the dispersionless lens is normalized to $-k_0 R/\lambda_0$ and is plotted in Supporting Information Fig. \ref{fig:S2_MaxDisps}(b) as a function of NA.

\section*{S6. Fermat's principle and the phase dispersion relation.} 
Phase only diffractive devices can be characterized by a local grating momentum (or equivalently phase gradient) resulting in a local deflection angle at each point on their surface. Here we consider the case of a 1D element with a given local phase gradient (i.e. $\phi_x=\partial\phi/\partial x$) and use Fermat's principle to connect the frequency derivative of the local deflection angle (i.e. chromatic dispersion) to the frequency derivative of $\phi_x$ (i.e. $\partial\phi_x/\partial\omega$). For simplicity, we assume that the illumination is close to normal, and that the element phase does not depend on the illumination angle (which is in general correct in local metasurfaces and diffractive devices).
Considering Fig. \ref{fig:S16_Fermat}(a), we can write the phase acquired by a ray going from point A to point B, and passing the interface at $\mathrm{x}$ as:
\begin{equation}
\Phi(x,\omega) = \frac{\omega}{\mathrm{c}}[\mathrm{n_1}\sqrt{x^2+{y_A}^2}+\mathrm{n_2}\sqrt{{(d-x)}^2+{y_B}^2}]+\phi(x,\omega)
\label{eq:FermatPhaseOmega}\\
\end{equation}
\noindent{To minimize this phase we need:}
\begin{equation}
\frac{\partial\Phi(x,\omega)}{\partial x} = \frac{\omega}{\mathrm{c}}[\frac{\mathrm{n_1}x}{\sqrt{x^2+{y_A}^2}}+\frac{\mathrm{n_2}(d-x)}{\sqrt{{(d-x)}^2+{y_B}^2}}]+\phi_x=0.
\label{eq:FermatPhaseDiffOmega}\\
\end{equation}
\noindent{For this minimum to occur at point O (i.e. $x=0$)}:
\begin{equation}
\phi_x(\omega)=\frac{\omega}{\mathrm{c}}\frac{\mathrm{n_2}d}{r}=\frac{\mathrm{n_2}\omega}{\mathrm{c}}\sin(\theta(\omega))
\label{eq:FermatPhaseDiffOmegaAt0}\\
\end{equation}
\noindent{which is a simple case of the diffraction equation, and where $r=\sqrt{d^2+{y_B}^2}$ is the OB length. At $\omega+\mathrm{d}\omega$, we get the following phase for the path from A to B' [Fig. \ref{fig:S16_Fermat}(b)]:}
\begin{equation}
\begin{split}
\Phi(x,\omega+\mathrm{d}\omega) = & \frac{\omega+\mathrm{d}\omega}{\mathrm{c}}[\mathrm{n_1}\sqrt{x^2+{y_A}^2}\\
& +\mathrm{n_2}\sqrt{{(d-x+\mathrm{d}x)}^2+{(y_B+\mathrm{d}y)}^2} ]+\phi(x,\omega+\mathrm{d}\omega)
\end{split}
\label{eq:FermatPhaseOmegadOmega}
\end{equation}
\noindent{where we have chosen B' such that OB and OB' have equal lengths. Minimizing the path passing through O:}\\
\begin{equation}
\phi_x(\omega+\mathrm{d}\omega)=\frac{\omega+\mathrm{d}\omega}{\mathrm{c}}\frac{\mathrm{n_2}(d+dx)}{r}=\frac{\mathrm{n_2}(\omega+\mathrm{d}\omega)}{\mathrm{c}}\sin(\theta(\omega+\mathrm{d}\omega))
\label{eq:FermatPhaseDiffOmegadOmegaAt0}\\
\end{equation}
\noindent{subtracting \ref{eq:FermatPhaseDiffOmegaAt0} from \ref{eq:FermatPhaseDiffOmegadOmegaAt0}, and setting $\phi_x(\omega+\mathrm{d}\omega)-\phi_x(\omega)=\frac{\partial\phi_x}{\partial\omega}\mathrm{d}\omega$, we get:}
\begin{equation}
\frac{\partial\phi_x}{\partial\omega}=\frac{\mathrm{n_2}}{c}\sin(\theta(\omega))+\frac{\mathrm{d}\theta}{\mathrm{d}\omega}\frac{\mathrm{n_2}\omega}{c}\cos(\theta(\omega)).
\label{eq:FermatPhaseDisp}\\
\end{equation}
\noindent{One can easily recognize the similarity between \ref{eq:FermatPhaseDisp} and \ref{eq:disp_disp_phase}.}

\section*{S7. Relation between dispersion and quality factor of highly reflective or transmissive meta-atoms.} 
Here we show that the phase dispersion of a meta-atom is linearly proportional to the stored optical energy in the meta-atoms, or equivalently, to the quality factor of the resonances supported by the mata-atoms. To relate the phase dispersion of transmissive or reflective meta-atoms to the stored optical energy, we follow an approach similar to the one taken in chapter 8 of~\cite{Harrington2001} for finding the dispersion of a single port microwave circuit. We start from the frequency domain Maxwell's equations:
\begin{align}
\begin{split}
&\nabla \times E = i \omega \mu H, \\
&\nabla \times H = -i \omega \epsilon E,
\end{split}
\label{align:Maxwell}
\end{align}
and take the derivative of the Eq. \ref{align:Maxwell} with respect to frequency:
\begin{equation}
\nabla \times \frac{\partial E}{\partial \omega} = i \mu H +i \omega \mu \frac{\partial H}{\partial \omega},
\label{eq:Deriv_1}\\
\end{equation}
\begin{equation}
\nabla \times \frac{\partial H}{\partial \omega} = -i \epsilon E -i \omega \epsilon \frac{\partial E}{\partial \omega}.
\label{eq:Deriv_2}\\
\end{equation}
Multiplying Eq. \ref{eq:Deriv_1} by $H^*$ and the conjugate of Eq. \ref{eq:Deriv_2} by $\partial E/\partial \omega$, and subtracting the two, we obtain
\begin{equation}
\nabla \cdot  (\frac{\partial E}{\partial \omega} \times H^*)  = i \mu |H|^2 +i \omega \mu \frac{\partial H}{\partial \omega} \cdot H^* -i \omega \epsilon \frac{\partial E}{\partial \omega} \cdot E^*.
\label{eq:Div_1}\\
\end{equation}
Similarly, multiplying Eq. \ref{eq:Deriv_2} by $E^*$ and the conjugate of Eq. \ref{eq:Deriv_1} by $\partial H/\partial \omega$, and subtracting the two we find:
\begin{equation}
\nabla \cdot  (\frac{\partial H}{\partial \omega} \times E^*)  = -i \epsilon |E|^2 -i \omega \epsilon \frac{\partial E}{\partial \omega} \cdot E^* +i \omega \mu \frac{\partial H}{\partial \omega} \cdot H^*.
\label{eq:Div_2}\\
\end{equation}
Subtracting Eq. \ref{eq:Div_2} from Eq. \ref{eq:Div_1} we get:
\begin{equation}
\nabla \cdot  (\frac{\partial E}{\partial \omega} \times H^*-\frac{\partial H}{\partial \omega} \times E^*)  = i \mu |H|^2 +i \epsilon |E|^2.
\label{eq:Diff}\\
\end{equation}
Integrating both sides of Eq. \ref{eq:Diff}, and using the divergence theorem to convert the left side to a surface integral leads to:
\begin{equation}
\sideset{}{_{\partial V} }\oint(\frac{\partial E}{\partial \omega} \times H^*-\frac{\partial H}{\partial \omega} \times E^*)  = i\sideset{}{_V}\int(\mu |H|^2 + \epsilon |E|^2)dv =2iU,
\label{eq:Integral}\\
\end{equation}
where $U$ is the total electromagnetic energy inside the volume $V$, and $\partial V$ denotes the surrounding surface of the volume. Now we consider a metasurface composed of a subwavelength periodic array of meta-atoms as  shown in Fig. \ref{fig:S17_PlanarStruct}. We also consider two virtual planar boundaries $\Gamma_1$ and $\Gamma_2$ on both sides on the metasurface (shown with dashed lines in  Fig. \ref{fig:S17_PlanarStruct}). The two virtual boundaries are considered far enough from the metasurface that the metasurface evanescent fields die off before reaching them. Because the metasurface is periodic with a subwavelength period and preserves polarization, we can write the transmitted and reflected fields at the virtual boundaries in terms of only one transmission $t$ and reflection $r$ coefficients. The fields at these two boundaries are given by:
\begin{align}
\begin{split}
E_1&=E+rE \\
H_1&=-\hat{z}\times(\frac{E}{\eta_1}-r\frac{E}{\eta_1}) \\
E_2&=tE \\
H_2&=-t\hat{z}\times\frac{E}{\eta_2}
\end{split}
\label{align:Fields}
\end{align}
where $E$ is the input field, $E_1$ and $E_2$ are the total electric fields at $\Gamma_1$ and $\Gamma_2$, respectively, and $\eta_1$ and $\eta_2$ are wave impedances in the materials on the top  and bottom of the metasurface.

Inserting fields from Eq. \ref{align:Fields} to Eq. \ref{eq:Integral}, and using the uniformity of the fields to perform the integration over one unit of area, we get:
\begin{equation}
\frac{\partial r}{\partial \omega}r^*\frac{|E|^2}{\eta_1}+\frac{\partial t}{\partial \omega}t^*\frac{|E|^2}{\eta_2} =i\tilde{U}
\label{eq:Total_Deriv}\\
\end{equation}
where $\tilde{U}$ is the optical energy per unit area that is stored in the metasurface layer. For a loss-less metasurface that is totally reflective (i.e. $t=0$ and $r= e^{i\phi}$), we obtain:
\begin{equation}
\frac{\partial \phi}{\partial \omega} = \frac{\tilde{U}}{P_\mathrm{in}},
\label{eq:Freq_disp}\\
\end{equation}
where we have used $P_\mathrm{in}=|E|^2/\eta_1$ to denote the per unit area input power. Finally, the dispersion can be expressed as:
\begin{equation}
\frac{\partial \phi}{\partial \lambda} = \frac{\partial \phi}{\partial \omega} \frac{\partial \omega}{\partial \lambda}=-\frac{\omega}{\lambda}\frac{\tilde{U}}{P_\mathrm{in}}.
\label{eq:Wavelength_disp}\\
\end{equation}

We used Eq. \ref{eq:Wavelength_disp} throughout the work to calculate the dispersion from solution of the electric and magnetic fields at a single wavelength, which reduced simulation time by a factor of two. In addition, in steady state the input and output powers are equal $P_\mathrm{out}=P_\mathrm{in}$, and therefore we have:
\begin{equation}
\frac{\partial \phi}{\partial \lambda} = -\frac{1}{\lambda}\frac{\omega\tilde{U}}{P_\mathrm{out}}=-\frac{Q}{\lambda}
\label{eq:Disp_Q}\\
\end{equation}
where we have assumed that almost all of the stored energy is in one single resonant mode, and $Q$ is the quality factor of that mode. Therefore, in order to achieve large dispersion values, resonant modes with high quality factors are necessary.

\bibliographystyle{naturemag_noURL}
\bibliography{MetasurfaceLibrary}

\end{document}